\documentclass{amsart}
\usepackage[latin1]{inputenc}

\usepackage{amssymb}
\usepackage{amsmath,enumerate,rotate}
\usepackage{amssymb,latexsym}
\usepackage{epsfig}
\usepackage{graphicx}
\usepackage{natbib}
\usepackage{color}

\newtheorem{definition}{Definition}[section]
\newtheorem{theorem}[definition]{Theorem}
\newtheorem{lemma}[definition]{Lemma}
\newtheorem{proposition}[definition]{Proposition}

\def \vzs {v_0^*}
\def \bvzs {\bar v_0^*}
\def \betamax { \beta_{(\nu_t)}}

\def \trgal  {\hat g_\a}
\def \galpha { g_\a}

\newcommand{\bfc}{{\bf c}}
\newcommand{\qn}{\hat{q}_n}
\newcommand{\qj}{\hat{q}_j}
\newcommand{\qnj}{\hat{q}_{n-j}}

\def \CR{\mathcal{R}}

\def\CF { \mathcal{F}}

\def\CM { \mathcal{M}}

\def\CB {\mathcal{B}}
\def\CP { \mathcal{P}}

\def\CP {\mathcal{P}}
\def\CU {\mathcal{U}}

\def\b {\beta}
\def\a {\alpha}

\def\d {\delta}
\def\th {\theta}
\def\s {\sigma}
\def\r {\rho}

\def\eps {\epsilon}
\def\n {\nu}

\def\a {\alpha}

\def\J {\mathbb{I}}

\def\N {\mathbb{N}}

\def\G {\mathbb{G}}

\def\RE {\mathbb{R}}

\def\N {\mathbb{N}}
\def\P {\mathbb{P}}

\def \fine{\diamondsuit}

\renewcommand{\Re}{\mathbb{R}}

\usepackage{verbatim}

\begin{document}

\title[Speed of approach to equilibrium for an inelastic Kac model]{Probabilistic study of the speed of approach to equilibrium for an inelastic Kac model}


\thanks{Research partially supported by Ministero dell'Istruzione, dell'Universit\`a e della Ricerca
 (MIUR grant 2006/134526)}

\author{Federico Bassetti}%
\author{Lucia Ladelli}
\author{Eugenio Regazzini}

 \address{Universit\`a degli Studi di  Pavia, Dipartimento di Matematica, via Ferrata 1,
 27100 Pavia, Italy}
%
% \author{Lucia Ladelli}%
%
\address{Politecnico di Milano, Dipartimento di Matematica, P.zza Leonardo da Vinci 32,
 2133 Milano, Italy}
% \author{Eugenio Regazzini  }
 %\ead{bassetti@dimat.unipv.it}

 \address{Universit\`a degli Studi di  Pavia, Dipartimento di Matematica, via Ferrata 1,
 27100 Pavia, Italy}

% \author{Eugenio Regazzini \corauthref{cor}}
%\corauth[cor]{Corresponding author}
% \ead{eugenio@dimat.unipv.it}

% \address{Universit\`a degli Studi di  Pavia, Dipartimento di Matematica, via Ferrata 1,
% 27100 Pavia, Italy}

 %\corauth[]{}


\email{federico.bassetti@unipv.it} 
\email{lucia.ladelli@polimi.it}
\email{eugenio.regazzini@unipv.it}

%\address{Univerista di Pavia....}

\begin{abstract}
 {This paper deals with a one--dimensional model for granular materials, which boils down 
 to an inelastic version of the Kac kinetic equation, with inelasticity  parameter $p>0$. 
 In particular, the paper provides bounds for certain distances -- such as specific weighted $\chi$--distances
and the Kolmogorov distance -- between the solution of that equation and the limit. It is assumed that the even part 
of the initial datum (which determines the asymptotic properties of the solution) belongs to the domain of normal attraction 
of a symmetric stable distribution with characteristic exponent $\a=2/(1+p)$. 
With such initial data, it turns out that the limit 
exists and is just the aforementioned stable distribution. A necessary condition
for the relaxation to equilibrium is also proved. Some bounds are obtained without introducing any extra--condition. Sharper bounds, 
of an exponential type, are exhibited in the presence of additional assumptions concerning either the behaviour, near to the origin, 
of the initial characteristic function, or the behaviour, at infinity, of the initial probability distribution
function. 
%All the methods used throughout the paper are essentially inspired to previous work of Harald Cram\'er and its developments 
%due to Peter Hall. In fact, our results on the solution of the inelastic equation heavily depend on a deep study 
%of the convergence in distribution of certain sums of weighted independent random variables. 
  }

\end{abstract}


% AMS classification 60f05,82C40
%\end{frontmatter}

\maketitle

\section{Introduction}\label{s1}
This work deals with a one--dimensional inelastic kinetic model,
introduced in \cite{PulvirentiToscani2004},
that can be thought of as a generalization of the  Boltzmann-like 
equation due to Kac (\cite{Kac1956}). Motivations for 
research into equations for inelastic interactions can be found in many papers, generally
devoted to Maxwellian molecules. Among them, in addition to the already mentioned Pulvirenti and Toscani's 
paper, it is worth quoting: 
\cite{BobylevCarrilloGamba2000}, \cite{CarrilloCercignaniaGamba2000},
\cite{BobylevCercignani2002a,BobylevCercignani2002b,BobylevCercignani2002c,BobylevCercignani2003},
\cite{ErnstBrito2002}, \cite{BobylevCercignaniToscani2003}, \cite{BolleyCarrillo2007}.
See, in particular, the short but useful review in \cite{Villani2006}.
Returning to the main subject of this paper, the one-dimensional inelastic model we want to study reduces to the equation
 \begin{equation}\label{eq1}
  \left\{ \begin{aligned}
           &\frac{\partial}{\partial t}f(v,t)+f(v,t)=
\frac{1}{2\pi}\int_{\Re\times[0,2\pi)}\frac{
             \{f(v c(\th)
              -w s(\th),t)f(vs(\th)+w c(\theta),t)}{c(\th)^2+s(\th)^2}dwd\th\\
             %&\qquad \qquad\qquad \qquad\qquad \qquad -f(v,t)f(w,t)\}dwd\theta\\
           &f(v,0):=f_0(v)\qquad(t>0, v\in\Re)\\
          \end{aligned}
  \right.
 \end{equation}
where $f(\cdot,t)$ stands for the probability density function of the
velocity of a molecule at time $t$ and 
\[
c(\th):=\cos\theta|\cos\theta|^{p}, \qquad 
s(\th):=\sin\theta|\sin\theta|^{p}
\]
$p$ being a nonnegative parameter. When $p=0$, (\ref{eq1}) becomes
the Kac equation. 
It is easy to check that the Fourier transform $\phi(\cdot,t)$ of
$f(\cdot,t)$
satisfies equation
 \begin{equation}\label{eq2}
  \left\{ \begin{aligned}
           &\frac{\partial}{\partial t}
\phi(\xi,t)=\frac{1}{2\pi}\int_0^{2\pi}
             \phi(\xi s(\th),t)\phi(\xi c(\th),t)d\theta-\phi(\xi,t)\\
           &\phi(\xi,0):=\phi_0(\xi)\qquad(t>0, \xi\in\Re)\\
          \end{aligned}
  \right.
 \end{equation}
where $\phi_0$ stands for the Fourier transform of $f_0$. 

Equation (\ref{eq2}) can be considered independently of (\ref{eq1}),
thinking of $\phi(\cdot,t)$, for $t \geq 0$, as Fourier--Stieltjes
transform of a probability measure $\mu(\cdot,t)$,
with $\mu(\cdot,0):=\mu_0(\cdot)$. In this case, differently from
(\ref{eq1}), $\mu$ needn't be absolutely continuous, i.e. it needn't
have a density function with respect to the Lebesgue measure.

Following \cite{Wild1951}, $\phi$ can be expressed as 
 \begin{equation}\label{eq3}
  \phi(\xi,t)=\sum_{n\geq1}e^{-t}(1-e^{-t})^{n-1}\qn(\xi;\phi_0)\qquad(t\geq0, \xi \in \RE )
 \end{equation}
where %$\qn$ can be found by recursion as
 \begin{equation}\label{eq4}
\left \{
\begin{array}{ll}
\hat q_1(\xi,\phi_0):=\phi_0(\xi)& \quad \\
  \qn(\xi;\phi_0):=\frac{1}{n-1}\sum_{j=1}^{n-1}\qnj(\xi;\phi_0)\circ\qj(\xi;\phi_0)
  & \qquad(n=2,3,\dots)\\
\end{array}
\right .
 \end{equation}
and 
 \begin{equation*}
  g_1\circ g_2(\xi)=\frac{1}{2\pi}\int_0^{2\pi}
g_1(\xi c(\th))g_2(\xi s(\th))d\theta\qquad(\xi\in\Re)
 \end{equation*}
is the so--called {\it Wild product}. The Wild representation 
(\ref{eq3}) can be used to prove that the Kac equations (\ref{eq1})
and (\ref{eq2}) have a unique solution in the class of all absolutely
continuous
probability measures and, respectively, in the class of the
Fourier--Stieltjes transforms of {\it all} probability measures on 
$(\RE,\CB(\RE))$. Moreover, this very same representation, as pointed
out by \cite{McKean1966}, can be reformulated in such a way to show
that $\phi(\cdot,t)$ is the characteristic function of a completely
specified sum of real--valued random variables. 
This represents an important point for the methodological side of the
present work, consisting in studying significant asymptotic properties
of $\phi(\cdot,t)$, as $t \to +\infty$. Indeed, thanks to the McKean
interpretation, our study will take advantage of methods and results
pertaining
to the {\it central limit theorem} of probability theory. 

As to the organization of the paper, in the second part of the present section we provide the reader with preliminary
information -- mainly of a  probabilistic nature -- that is necessary to understand the rest of the paper.
In Section \ref{s2} we present the new results, together with a few hints to the strategies used to prove them. 
The most significant steps of the proofs are contained in Section
\ref{s3}, devoted to asymptotics for weighted sums of independent random variables. 
The methods used in this section are essentially inspired to previous work of Harald Cram\'er 
and to its developments due to Peter Hall. See \cite{Cramer1962,Cramer1963}, \cite{Hall1981}.
Completion of the proofs 
is deferred to the Appendix.

\subsection{Probabilistic interpretation of solutions of (\ref{eq1})--(\ref{eq2})}
\begin{figure}
        \includegraphics [scale=0.5]{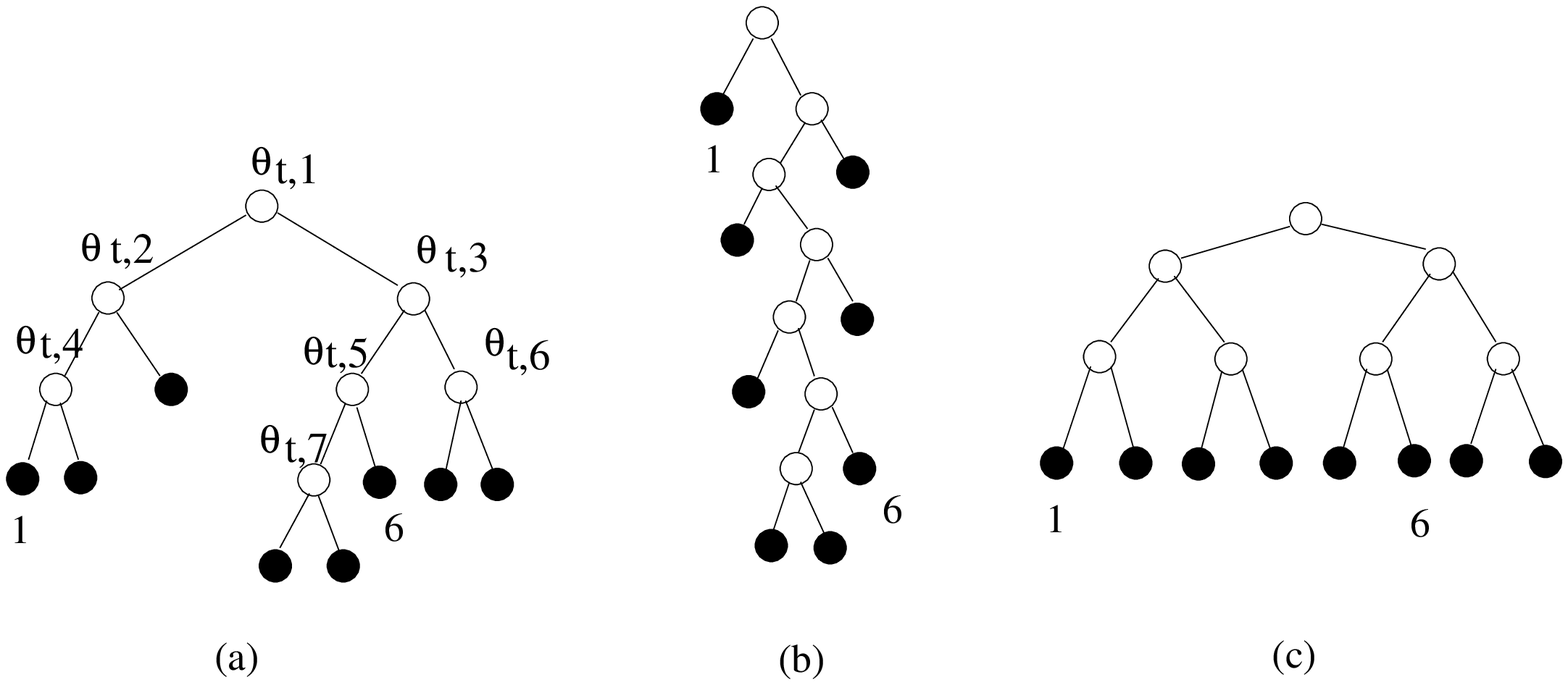}
   \caption{}\label{figure1}
\end{figure}

It is worth lingering over the McKean reformulation of (\ref{eq4}),
following \cite{GabettaRegazziniCLT}. Consider the product spaces
\[
\Omega_t:=\N \times \G \times [0,2\pi)^{\N}\times \RE^{\N}
\]
with $\G=\bigcup_n G(n)$, $G(n)$ being a set of certain {\it binary
  trees}
with $n$ leaves. These trees are defined so that each node has either
  zero
or two ``children'', a ``left child'' and a ``right child''. See
  Figure \ref{figure1}.

Now, equip $\Omega_t$ with the $\s$--algebra 
\[
\CF_t:=\CP(\N\times \G) \otimes \CB([0,2\pi)^{\N}) \otimes \CB(\RE^\N)
\]
where, given any set $S$, $\CP(S)$ denotes the power set of $S$ and, 
if $S$ is a topological space, $\CB(S)$ indicates the Borel
$\s$--algebra on $S$. Define
\(
(\nu_t, \gamma_t,\th_t, X_t) \), with  $\th_t:=(\th_{t,n})_{n \geq 1}$ and  $X_t:=(X_{t,n})_{n \geq 1}$,
to be the coordinate random variables of $\Omega_t$. At this stage, for
each tree in $G(n)$ fix an order on the set of all the $(n-1)$ nodes
and, accordingly, associate the random variable $\th_{t,k}$ with the
$k$--th node. See (a) in Figure \ref{figure1}. 
Moreover, call $1,2,\dots,n$ the $n$ leaves following a left to right order. See
(b) in Figure \ref{figure1}.
Define the depth of leaf $j$ -- in symbols, $\d_j$ -- to be the number
of generations which separate $j$ from the ``root'' node, and for each
leaf $j$ of a tree, form the product 
\[
\beta_{j,t}:=\prod_{i=1}^{\d_j} \a_i^{(j)}
\]
where: $\a_{\d_j}^{(j)}$ equals $c(\th_{t,k})$ if $j$ is a ``left
child'' or  $s(\th_{t,k})$ if $j$ is a ``right
child'', and $\th_{t,k}$ is  the element of $\th_t$ associated to the
parent node of $j$; $\a_{\d_j-1}^{(j)}$ equals $c(\th_{t,m})$
or  $s(\th_{t,m})$ depending on the parent of $j$ is, in its turn, 
 a ``left child'' or  a ``right child'', 
 $\th_{t,m}$ being the element of $\th_{t}$  associated with the
grandparent of $j$; and so on. For the unique tree in $G(1)$ it is
assumed that $\beta_{1,t}=1$.
For instance, as to leaf 1 in (a) of  Figure  \ref{figure1},
$\beta_{1,t}=c(\theta_{t,4})\cdot c({\theta}_{t,2})\cdot
c({\theta}_{t,1})$
and, for leaf 6,
$\beta_{6,t}=s ({\theta}_{t,5})\cdot
c({\theta}_{t,3}) \cdot s({\theta}_{t,1})$.

From the definition of the random variables $\beta_{j,t}$ it is plain to deduce that
\[
\sum_{j=1}^{\n_t}|\beta_{j,t}|^\a=1,
\]
holds true for any tree in $G(\nu_t)$, with
\[
\a:=\frac{2}{1+p},
\]

For further information on this construction, see 
\cite{McKean1967,CarlenCarvalhoGabetta2000,GabettaRegazziniCLT}.

It is easy to verify that there is one and only one probability
measure $P_t$ on $(\Omega_t,\CF_t)$ such that
\[
\begin{split}
P_t\{ &\nu_t=n,\gamma_t=g,\th_t \in A, X_t \in B  \} \\
&=\left \{
\begin{array}{cc}
e^{-t}(1-e^{-t})^{n-1} p_n(g) u^{\otimes \N}(A) \mu_0^{\otimes \N}(B)
& \text{if $g \in G(n)$}\\
0 &  \text{if $g \not \in G(n)$} \\
\end{array}
\right .
\end{split}
\]
where, for each $t$, 
\begin{itemize}
\item $p_n$ is a well--specified probability on $G(n)$, for every $n$.
\item $u^{\otimes \N}$ is the probability distribution that makes the
  $\th_{t,n}$ independent and identically distributed with continuous
  uniform law on $[0,2\pi)$.
\item $\mu_0^{\otimes \N}$ is the probability distribution according
  to which the random variables $X_{t,n}$ turn out to be independent and identically
  distributed with common law $\mu_0$.
\end{itemize}

Expectation with respect to $P_t$ will be denoted by $E_t$ and integrals over a measurable 
set $A \subset \Omega$ will be often indicated by $E_t(\cdot;A)$.

In this framework one gets the following proposition, a proof of which
can be obtained from obvious modifications of the proofs of Theorem 3
and Lemma 1 in \cite{GabettaRegazziniCLT}. 

($F_1$)   {\it  The solution $f(\cdot,t)$  {\rm[}$\phi(\cdot,t)$, 
respectively {\rm]}
of  {\rm(\ref{eq1})}  {\rm[(\ref{eq2})}, respectively {\rm]} can be viewed as a
probability density function  {\rm[}the characteristic function,
  respectively {\rm]}
of 
\[
V_t:=\sum_{j=1}^{\nu_t} \beta_{j,t} X_{t,j}
\]
for any $t >0$.
Moreover, $\b_{(\nu_t)}:=\max\{|\beta_{1,t}|,\dots,|\beta_{\nu_t,t}| \}$ converges in distribution 
to zero as $t \to +\infty$.
}

As a first application of this proposition, one easily gets
\[
\begin{split}
\phi(\xi,t)&=E_t[E_t(e^{i\xi V_t}|\nu_t)]\\
           &=e^{-t}\phi_0(\xi)+e^{-t}\sum_{n \geq 2} (1-e^{-t})^{n-1}
\qn(\xi;\phi_0).
\end{split}
\]
Then, since $\qn(\xi;\phi_0)=\qn(\xi;Re(\phi_0))$
for  any $n \geq 2$ --- with $Re(z)=$real part of $z$ --- the conditional
characteristic function of $V_t$, given $\{\nu_t=n\}$,
coincides with the characteristic function of $V_t$ when $\phi_0$ is
replaced by its real part. Whence, 
\begin{equation}\label{eq5bis}
\phi(\xi,t)=e^{-t}\sum_{n \geq 1}  (1-e^{-t})^{n-1} \qn(\xi;Re(\phi_0))+i
Im(\phi_0(\xi)) e^{-t}
\end{equation}
with $Im(z):=$imaginary part of $z$. The distribution corresponding to
$Re(\phi_0)$ is symmetric and is called {\it even part}
of $\mu_0$. In fact, $Re(\phi_0)$ turns out to be an even 
real--valued characteristic function, and this fact generally
makes easier certain computations. It should be pointed out that if the
initial
datum $\mu_0$ is a symmetric probability distribution, then the 
distribution of $V_t$ is the same as the distribution 
of $\sum_{j=1}^{\nu_t}|\beta_{j,t}| X_{t,j}$.

\subsection{Topics on stable distributions}\label{Subec_2.1}
It can be proved that the possible limits (in distribution)
of $V_t$, as $t \to +\infty$, have characteristic functions $\phi$ which 
are solutions of 
\begin{equation}\label{eq6}
\frac{1}{2\pi}\int_{0}^{2\pi} \phi(\xi s(\th))\phi(\xi
c(\th))d\th=\phi(\xi)
\qquad (\xi \in \RE).
\end{equation}
This result has been communicated to us by Filippo Riccardi, who
proved it by resorting to a suitable modification of the Skorokhod
representation used in the Appendix of the present paper. It is
interesting to note that also the stationary solutions of (\ref{eq2}) must
satisfy (\ref{eq6}). We didn't succeed in finding all the solutions of
(\ref{eq6}), but it is easy to check that 
\begin{equation}\label{eq7}
 \trgal(\xi)=\exp\{-a_0 |\xi|^\a \} \qquad  (\xi \in \RE)
\end{equation}
is a solution of  (\ref{eq6}), for any $a_0\geq 0$. 

It is well--known that (\ref{eq7})
is strictly connected with certain sums of random variables. Indeed,
it is a stable real--valued characteristic function with
characteristic exponent $\a$ and, in view of a classical L\'evy's
theorem, 

($F_2$)   {\it If $X_1,X_2,\dots$ are independent and identically
  distributed real--valued random variables, with symmetric common
  distribution 
function $F_0$, then in order that the random variable $X$ be the
  limit in distribution of the normed sum $\sum_{i=1}^n X_i/n^{1/\a}$
it is necessary and sufficient that $X$ has characteristic function
  {\rm (\ref{eq7})}
for some $a_0 \geq 0$. }

One could guess that ($F_2$) may be used to get a direct proof of the
fact that $V_t$ converges in distribution to a stable random variable
with characteristic function  (\ref{eq7}). This way, one would obtain
that these characteristic functions are all possible pointwise limits,
as $t \to +\infty$, of solutions $\phi(\cdot,t)$ of (\ref{eq2}). In
point of fact, direct application of results like ($F_2$) is
inadmissible since $V_t$ is a weighted sum of a random number of
summands, affected by 
random weights which are not stochastically independent. In spite of
this, by resorting to suitable forms of conditioning for $V_t$, one
can take advantage of classical propositions pertaining to the
 central limit theorem. 

In addition to the problem of determining the class of all possible
limit distributions for $V_t$, an obvious question which arises is
that of singling out  necessary and sufficient conditions on $\mu_0$, 
in order that $V_t$ converges in distribution to some specific
random variable. As to the classical setting mentioned in ($F_2$), it
is worth recalling 

($F_3$)   {\it  If $X_1,X_2,\dots$ are independent and identically distributed
real--valued random variables, with {\rm(}not necessarily symmetric{\rm)} common distribution
function $F_0$, then in order that 
$(\sum_{i=1}^n X_i/n^{1/\a}-m_n)$ converge in law to a random variable with
characteristic function  {\rm(\ref{eq7})}  with some specific value for
$a_0$ --- or, in other words, that $F_0$ belong to the  domain of normal
attraction of   {\rm(\ref{eq7})} --- it is necessary
and sufficient that $F_0$ satisfies $|x|^\a F_0(x) \to c_1$ as $x \to -\infty$ and
$x^{\a}(1-F_0(x)) \to c_2$ as $x \to +\infty$, i.e.}
\begin{equation}\label{NDA-F}
\begin{split}
F_0(-x)&=\frac{c_1}{|x|^\a}+S_1(-x)  \quad \text{\it and} \quad  1-F_0(x)=\frac{c_2}{x^\a} + S_2(x) \qquad (x>0) \\
S_i(x)&=o(|x|^{-\a}) \quad \text{\it as} \quad |x| \to +\infty \quad (i=1,2).\\
\end{split}
\end{equation}

For more information on stable laws and central limit theorem see, 
for example, Chapter 2 of \cite{IbragimovLinnik1971} and 
Chapter 6 of \cite{Galambos1995}. To complete the description of certain 
facts that will be mentioned throughout the paper, it is worth
enunciating

($F_4$) {\it If $\phi_0$ stands for the Fourier--Stieltjes transform
  of a probability distribution function $F_0$ satisfying 
{\rm(\ref{NDA-F})}, then 
\begin{equation*}%\label{NDA-F3}
1-\phi_0(\xi)=(a_0+   v_0(\xi))|\xi|^\a  \qquad (\xi \in \RE)
\end{equation*}
where $v_0$ is bounded and $|v_0(\xi)|=o(1)$ as $|\xi|\to 0$. }

($F_4$), which is a paraphrase of Th\'eor\`eme 1.3 of
\cite{Ibragimov1985}, can be proved by mimicking the argument used for
Theorem 2.6.5 of \cite{IbragimovLinnik1971}.

\section{Presentation of the new results}\label{s2}

In the present paper our aims are: Firstly, to find initial distribution functions $F_0$ (or initial 
characteristic functions $\phi_0$) so that the respective solutions of (\ref{eq2})
may converge pointwise to (\ref{eq7}). Secondly, to determine the rate of convergence of the probability 
distribution function $F(\cdot,t)$, corresponding to $\phi(\cdot,t)$, to a stable distribution 
function $G_\a$ with characteristic exponent $\a=2/(1+p)$, with respect both to specific
 weighted $\chi$--metrics 
and to Kolmogorov's distance. 

It is well--known --- from the L\'evy continuity theorem --- that pointwise convergence of sequences of characteristic functions 
is equivalent to {\it weak convergence} of the corresponding distribution functions. In particular, in our present case, 
since the limiting distribution function $G_\a$ is (absolutely) continuous, weak convergence is equivalent to uniform convergence, i.e. 
\begin{equation}\label{eq9}
\sup_{x \in \RE} |F(x,t)-G_\a(x)| \to 0 \qquad \text{as} \,\, t \to +\infty.
\end{equation}

Left--hand side of (\ref{eq9}) is just the {\it Kolmogorov distance} ($K$, in symbols)
between $F(\cdot,t)$ and $G_\a$. As to the above--mentioned first aim, besides sufficient conditions for convergence 
--- reducing to the fact that $F_0$ belongs to the domain of normal attraction of (\ref{eq7}) ---
a necessary condition for convergence is given. As far as rates of convergence are concerned, results can be found in the paper
of Pulvirenti and Toscani, with respect to a specific {\it weighted $\chi$--metric}, used to study convergence to equilibrium 
of Boltzmann--like equations starting from \cite{GabettaToscaniWennberg1995}. See also \cite{Rachev1991}.
Denoting this distance by $\chi_s$, $s$ being some positive number, one has
\begin{equation*}\label{eq10}
\chi_s(F(\cdot,t),G_\a) := \sup_{\xi \in \RE} \frac{|\phi(\xi,t)-\exp(-a_o|\xi|^\a)|}{|\xi|^s}.
\end{equation*}
With reference to (\ref{eq1}), after writing $\galpha$ for a density of $G_\a$, Theorem 6.2 
in \cite{PulvirentiToscani2004} reads:

($F_5$) {\it 
Let $p>1$ with $f_0$ such that  %$|v_0(\xi)|=O(|\xi|^\delta)$ as $\xi \to 0$
$\int_\RE |v|^{\a +\delta} |f_0(v)-\galpha(v)|dv$ is finite 
for some $\delta$ in $(0,(1-\a)\wedge  \a)$. Then
\begin{equation}\label{eq11}
\chi_{\a+\delta}(F(\cdot,t),G_\a) \leq \chi_{\a+\delta}(F_0,G_\a) \exp\{-t(1-2A_{2(1+\delta/\a)})\}
\end{equation}
holds true for every $t \geq 0$, with 
\begin{equation}\label{eq12}
A_m:=\frac{1}{2\pi} \int_0^{2\pi} |\sin \th|^m d\th=\frac{\Gamma(\frac{m}{2}+\frac{1}{2})}{\sqrt{\pi}\,\, \Gamma(\frac{m}{2}+1)} \qquad (m \geq 0).
\end{equation}
Moreover, {\rm(\ref{eq11})} is still valid 
if $0< p\leq 1$ and $\int_\RE|v|^{\a+\delta}|f_0(v)-\galpha(v)|dv$
if finite 
for some  $\delta$ in $(0,\a p]$.
}

It should be pointed out that the proof of ($F_5$) provided in
\cite{PulvirentiToscani2004} rests on a hypothesis that is weaker than the one evoked in ($F_5$),  
i.e.
\begin{equation}
  \label{eq:13bis}
  |v_0(\xi)|=O(|\xi|^\delta) \qquad \text{ as $\xi \to 0$}
\end{equation}
for some $\d>0$.

In the present paper we prove weak convergence of $F(\cdot,t)$ to $G_\a$ under much more general 
hypotheses than those
adopted in ($F_5$). For 
reader's convenience, it is worth noticing that the probability distribution function 
$F_0^*$ corresponding to $Re(\phi_0)$ (see the final part of Subsection \ref{Subec_2.1}) coincides with
\[
\frac{1}{2}\{F_0(x)+1-F_0(-x)\}
\]
at each point $x$ of continuity for $F_0$. In view of ($F_3$)--($F_4$), if $F_0$ belongs to the domain of normal 
attraction of (\ref{eq7}), then there is a nonnegative $c_0$ for which
\begin{equation}\label{eq12bis}
\lim_{x \to -\infty}|x|^\a F^*_0(x)=\lim_{x \to +\infty} x^\a (1-F^*_0(x))=c_0
\end{equation}
and the characteristic function associated to $F_0^*$, i.e. $Re(\phi_0)$, satisfies 
\begin{equation}\label{eq12ter}
1-Re(\phi_0(\xi))=(a_0+v_0^*(\xi))|\xi|^\a
\end{equation}
for some bounded, real-valued $v_0^*$ such that $|v_0^*(\xi)|=o(1)$ as $\xi \to 0$. Moreover, $c_0$ is related to $a_0$
by 
\[
a_0=2 c_0 \int_0^{+\infty} \frac{\sin(x)}{x^\a}dx.
\]

The precise statement of the aforementioned convergence reads

\begin{theorem}\label{Thm1} Given $p>0$, let 
the initial data for problems   {\rm(\ref{eq1})--(\ref{eq2})}  satisfy 
\[
\lim_{x \to +\infty} (1-F_0^*(x))x^{\a}=c_0.
\]
Then 
\[
\lim_{t \to +\infty} K(F(\cdot,t),G_\a)=0.
\]
In particular, if $c_0=0$, then for every $\eps>0$ one has
\[
\lim_{t \to +\infty}F(-\eps,t)=\lim_{t \to +\infty}(1-F(\eps,t))=0,
\]
i.e. the weak limit of $\mu(\cdot,t)$ is the point mass $\delta_0$. 
On the other hand, if $p>0$ and there is a strictly positive and increasing sequence 
$(t_n)_{n \geq 1}$, divergent to $+\infty$, such that $(F(\cdot,t_n))_{n \geq 1}$ converges weakly
to any probability distribution function, then
\[
0 \leq \lim_{\xi \to +\infty} \inf_{x \geq \xi} x^\a(1-F_0^*(x)) < +\infty.
\]
\end{theorem}

Proof of Theorem \ref{Thm1} is deferred to the Appendix.

After presenting the most general statement we achieved about the weak convergence of $F(\cdot,t)$, let us
proceed to investigate how convergence is fast. 
Pulvirenti and Toscani's argument to prove ($F_5$) lies in studying equation (\ref{eq4}) directly
via suitable inequalities and from an analytical viewpoint. Differently, in our approach one starts from 
inequality 
\begin{equation}\label{eq13}
|\phi(\xi,t)-\trgal(\xi)| \leq E_t(|\tilde \phi_{\nu_t}(\xi)-\trgal(\xi)|)
\end{equation}
where $\trgal$ is defined by (\ref{eq7}) and, according to ($F_1$), $\tilde \phi_{\nu_t}$
represents the conditional characteristic function of $V_t$ given 
$(\nu_t,\gamma_t,\th_t)$. Hence, from the beginning, we try to obtain bounds for 
$|\tilde \phi_{\nu_t}(\xi)-\trgal(\xi)|$. This is tantamount to investigating bounds for
$|\tilde \phi_{n}(\xi)-\trgal(\xi)|$ when $\tilde \phi_{n}$ is the characteristic function of 
\begin{equation}\label{eq14}
S_n:=\sum_{l=1}^n q_l^{(n)} X_l
\end{equation}
with $X_1,X_2,\dots$ independent and identically distributed random numbers, with
common distribution function $F_0$, and  
\begin{equation}\label{eq15_1}
q_l^{(n)} \geq 0 \,\, \,\, \text{for} \,\,\,\, l=1,\dots,n,\,\,\, \, n=1,2,\dots \,\,\, \text{such that} \,\,\, 
\sum_{l=1}^n (q_l^{(n)})^{\a}=1.
\end{equation}
Think of $n$ and $(q_1^{(n)},\dots,q_n^{(n)})$ as realizations of $\nu_t$
and $(|\beta_{1,t}|,\dots,|\beta_{\nu_t,t}| )$, respectively. According to ($F_1$) one can assume 
\begin{equation}\label{eq15_2}
q_{(n)}:=\max\{q_1^{(n)},\dots,q_n^{(n)}  \} \to 0 \quad \text{as $n \to +\infty$}.
\end{equation}
We study this problem -- preliminary to the investigation of rates of convergence for $V_t$ --
under the additional conditions that $F_0$ is {\it symmetric} 
(and, consequently, the corresponding
characteristic function $\phi_0$ is even) and that it belongs to the domain of normal attraction of $\trgal$. See
($F_3$)--($F_4$) and (\ref{eq12bis})--(\ref{eq12ter}). This way we also get bounds for convergence in law of weighted sums $S_n$ to stable random variables, which are of interest in themselves and, as far as we know, seem to be new. They are explained and precisely formulated in Section 3. 
Resuming now the main issue of the speed of convergence of $V_t$ to equilibrium, some further notation is needed. We set 
\[
\|\vzs\|:=\sup_{\xi \geq 0} |\vzs(\xi)|, \quad M:=a_0+\|\vzs \|, \quad 
\bvzs(\xi):=\sup_{0 \leq x  \leq \xi} |\vzs(x)|
\]
and, given $\eta \in (0,a_0)$, define $d$ to be some element of $(0,1)$ such that
\[
\frac{4}{5} M^2 |x|^\a +\bvzs(x) \leq \eta 
\]
comes true whenever $|x| \leq (3d/(8M))^{1/\a}$. Next, we put 
%\[
%D_n=D_n(c,d):= \left(\frac{3}{8M} \{d \wedge |x|^{c\a} \}  \right)^{1/\a}
%|x|^{-1}
%\]
$M_r:=\max_{x \geq 0} x^{r\a}e^{-(a_0-\eta)x^\a}$, 
$d_1:=(3/(8M))^{1/\a}$, $k^*=\bvzs(d_1 d^{1/\a})(1+2d_1^\a d^{1-\a}
\bvzs(d_1 d^{1/\a})) + (4/5) M^2 d_1^\a d +(32/25) M^4 d_1^{3\a} d^{3-\a}$.

\subsection{Speed of approach to equilibrium with respect to weighted $\chi$--metrics}\label{Subsec2.1}
Now we are in a position to present our first results which concern convergence of $F(\cdot,t)$
to $G_\a$ with respect to $\chi$--metrics.

\begin{theorem}\label{thm2}
Let $F_0$ belong to the domain of normal attraction of $G_\a$ with $\a=2/(1+p)$, for some $p>0$.
Define $v_0$ and $v_0^*$ to be the same as in {\rm ($F_4$)} and  {\rm (\ref{eq12ter})}, respectively. Set 
\(
\betamax:=\max\{|\beta_{1,t}|,\dots,|\beta_{\nu_t,t}| \}.
\) 
Then
\[
\begin{split}
\chi_\a(F(\cdot,t),G_\a) & \leq E_t(\bvzs(d_1 \betamax^c)) + 2M_1 E_t 
(\bvzs(d_1 \betamax^c)^2)
+\frac{4}{5} M^2 M_1 E_t(\betamax^\a ) \\ & +\frac{32}{25} M_3 M^4 E_t(\betamax^{2\a}) 
+ \big(k^*+\frac{2}{d d_1^\a}\big) P_t\{ \betamax > d \wedge d^{1/c\a}\}\\ &  + 
\frac{2}{d_1^\a} E_t(\betamax^{\a(1-c)}) + e^{-t} \sup_{\xi \in \RE} |Im( v_0 (\xi))|\\
\end{split}
\]
is valid for any $c$ in $(0,1)$. 
\end{theorem}

The upper bound provided in Theorem \ref{thm2} goes to zero as $t \to +\infty$ thanks 
to ($F_1$), ($F_4$) and the definition of $\bvzs$. Then, it
can be used to yield further bounds, either via the
statement of specific upper bounds for the expectations which appear in the right--hand side, or
through the adoption of suitable extra--conditions on $v_0$. As to the former way of arguing, it is worth
recalling that Proposition 8 in \cite{GabettaRegazzini2006a}
gives
\begin{equation}\label{eq16}
\begin{split}
E_t(\sum_{j=1}^{\nu_t}|\beta_{j,t}|^m )&=E_t( \sum_{j=1}^{\nu_t}A_{m(1+p)}^{\delta_j} )
\quad \text{($\d_j=$ depth of leaf $j$)} \\
&=\exp\{-t(1-2A_{m(1+p)}) \} \quad (m \geq 0) \\
\end{split}
\end{equation}
with $A_m$ defined as in (\ref{eq12}).
 Moreover, from Lemma 1 in \cite{GabettaRegazziniCLT}, 
\begin{equation}\label{eq17}
P_t\{ \betamax > x \} \leq x^{-\frac{q}{1+p}} e^{-t(1-2A_q)} \qquad (0<x<1, q>0)
\end{equation}
which, in turn, yields 
\begin{equation}\label{eq18}
 E_t(\betamax^m) \leq e^{-\sigma m t}+ e^{-t(1-q\sigma \a/2 -2A_q)}
\end{equation}
for any positive $\s$ and $q$. Now, define $\CU_{1,t}$ as
\[
\begin{split}
\CU_{1,t}:&=\bvzs(d_1 \betamax^c)+2M_1(\bvzs(d_1 \betamax^c))^2+\frac{4}{5}
M^2M_1\betamax^\a +\frac{32}{25} M_3 M^4 \betamax^{2\a} \\
&+(k^*+\frac{2}{d d_1^\a}) \J\{ \betamax > d \wedge d^{1/c\a}\} 
 + \frac{2}{d_1^\a} \betamax^{\a(1-c)} + e^{-t} \sup_{\xi \in \RE} |Im(v_0 (\xi))| \\
\end{split}
\]
and set
\[
\begin{split}
\CM_{1,t}&:=\bvzs(d_1 \betamax^c)+2M_1(\bvzs(d_1 \betamax^c))^2 \\
\CR_{1,t}&:=\CU_{1,t}-\CM_{1,t}. \\
\end{split}
\]
Next, observe that the upper bound provided by Theorem \ref{thm2}
can be written as
\[
E_t(\CM_{1,t})+E_t(\CR_{1,t}) \leq 
E_t(\CM_{1,t};\betamax \leq x_t)+M(1+2M_1M) P_t\{\betamax > x_t \} +E_t(\CR_{1,t})
\]
with $x_t:=\exp\{-\s t \}$ and $\s$ satisfying $1-2A_q-\s q/(1+p)>0$ to obtain
\begin{equation}\label{eq19}
\begin{split}
\chi_\a(F(\cdot,t),G_\a) \leq & \bvzs(d_1e^{-c\s t}) +2M_1 \bvzs(d_1e^{-c\s t})^2
\\ &+M(1+2M_1M) e^{-t(1-2A_q-\s q/(1+p))} + E_t(\CR_{1,t}). \\
\end{split}
\end{equation}
Then, since $E_t(\CR_{1,t})$ can be re-written --- thanks to (\ref{eq17})--(\ref{eq18}) ---
as a sum of exponential functions, (\ref{eq19}) provides a bound entirely expressed, through 
$\bvzs$, in terms of exponential functions of $t$.
%
%functions of that very same type. 

Exponential rates of relaxation to equilibrium hold true under some extra--condition concerning 
the local behavior of $v_0$ near the origin.

\begin{theorem}\label{thm3} Assume that, in addition to the assumptions made in 
{\rm Theorem \ref{thm2}}, %$|v_0(\xi )|=O(|\xi|^{\delta})$ as $\xi\rightarrow 0$ for some $\delta >0$
{\rm (\ref{eq:13bis})} holds  for some $\delta >0$. Moreover, let $d$ be chosen in such a
way that $|x| \leq d_1d^{1/\a}$ entails  $|v_0(x)| \leq \rho |x|^{\delta}$ for some $ \rho>0$. Then,
\[
\begin{split}
\chi_{\a + \d} \big(F(\cdot,t),G_\a \big) \leq & \Big(\r+\frac{2}{d_1^{\a+\d}d^{(\a+\d)/\a}}\Big) e^{-t(1-2A_{2(1+ \d/\a)})} \\
+ & \frac{4}{5}M^2M_{\frac{\a-\d}{\a}} e^{-t(1-2A_{4})}+2\r^2M_{\frac{\a +\d }{\a}}e^{-t(1-2A_{2(1+ 2\d/\a)})}\\
+& \frac{32}{25}M^4M_{\frac{3\a-\d}{\a}} e^{-t(1-2A_{6})}+ e^{-t} \sup_{\xi \in \RE} \frac{1}{|\xi|^{\d}}|Im( v_0 (\xi))| \\
\end{split}
\]
holds true for ${\delta}$ in $(0, \a]$, while
\[
\begin{split}
\chi_{2\a} \big (F(\cdot,t),G_\a \big) \leq & \Big(\frac{4}{5}M^2+\frac{2}{ d_1^{2\a}d^2}\Big) e^{-t(1-2A_{4})}\\
+ &\r M_{\frac{\d-\a}{\a}}e^{-t(1-2A_{2(1+ \d/\a)})}+\frac{32}{25}M^4M_2 e^{-t(1-2A_{6})}\\
+& 2\r^2M_{\frac{2\d}{\a}}e^{-t(1-2A_{2(1+ 2\d/\a)})}+ e^{-t} \sup_{\xi \in \RE} \frac{1}{|\xi|^{\a}}|Im( v_0 (\xi))| \\
\end{split}
\]
is verified for ${\delta}$ in $(\a, 2\a]$.
\end{theorem}

In short, this proposition can be condensed into the following statement: {\it Under the hypotheses of 
{\rm Theorem \ref{thm3}}, there are constants $a_1$ and $a_2$ such that:
\[
\begin{split}
\chi_{\a + \d} (F(\cdot,t),G_\a) \leq & a_1  e^{-t(1-2A_{2(1+ \d/\a)})} \qquad \text{if} \quad \d \in (0, \a],\\
\chi_{2\a} (F(\cdot,t),G_\a) \leq &  a_2  e^{-t(1-2A_{4})} \qquad \text{if} \quad  \d \in (\a, 2\a].\\
\end{split}
\]
}

Statements of the same type as Theorems \ref{thm2} and 
\ref{thm3} are proved in Section 5 in  
\cite{GabettaRegazziniCLTspeed}
for $\a=2$ $(p=0)$, i.e. when $G_\a$ is a Gaussian 
distribution function with zero mean. Notice that the rate of convergence 
given in the former part of the last theorem coincides with that of Toscani and Pulvirenti 
previously quoted in
($F_5$). The latter part of  Theorem \ref{thm3} and, mainly,  Theorem \ref{thm2} seem to be new. 
See Subsection \ref{Subsec2.4} for further comments.

\subsection{Rates of relaxation to equilibrium in Kolmogorov's metric (Conditions expressed on the 
characteristic function $\phi_0$).}\label{Subsec2.2}

Rates of convergence of $F(\cdot,t)$ to $G_\a$, in  Kolmogorov's metric, can be deduced from the 
representation of $V_t$ as weighted sum, via 
the well-known Berry-Esseen inequality in its form given, for example, in 
Theorem 3.18 of \cite{Galambos1995}. It is worth recalling that application 
of this inequality is allowed thanks to the fact that $G_\a$ has derivatives of all orders at every point. 
Henceforth, given any strictly positive 
$l$ and $q$, we put 
\[
N_l=\int_0^{+\infty}\exp\{-(a_0-\eta)\xi^{\a}\}\xi^{l-1}
d\xi
\] 
and 
\[
H(\xi,q):=|\vzs(\xi q)|(1+2|\xi|^{\a}|\vzs(\xi q)|), \qquad 
\bar{H}(\xi,q):=\sup_{u\leq q} H(\xi ,u)
\]
with $\vzs$ as in (\ref{eq12ter}).

\begin{theorem}\label{thm4}
If $F_0$ belongs to the domain of normal attraction of  $G_\a$ with $\a=2/(1+p)$ for some $p>0$, then 
\[
\begin{split}
K(F(\cdot,t),G_\a) \leq & \frac{2}{\pi}E_t\big[\sum_{j=1}^{\nu_t}|\beta_{j,t}|^{\a}\int_0^{+\infty}H(\xi,|\beta_{j,t}|)\xi^{\a-1} 
e^{-(a_0-\eta)\xi^{\a}}d\xi \big]\\
+& \frac{{\bfc}||h_{\a}||}{\tilde{d}}E_t(\betamax)+ \frac{8}{5\pi}M^2N_{2\a}e^{-t(1-2A_{4})} +  
\frac{64}{25\pi}M^4N_{4\a}e^{-t(1-2A_{6})}\\
+&  \frac{e^{-t}}{2}\sup_{x \in \RE}|F_0(x)+F_0(-x-0)-1|
\end{split}
\]
$\bfc$ being the constant which appears in the above-mentioned version of the
 Berry-Esseen inequality and $\tilde{d}:=\big(
{3d}/{8M}\big)^{1/\a}$.
\end{theorem}

A further bound for $K(F(\cdot,t),G_\a)$ can be obtained by replacing the summand  
\[
\frac{2}{\pi}E_t[\sum_{j=1}^{\nu_t}|\beta_{j,t}|^{\a}\int_0^{+\infty}H(\xi,|\beta_{j,t}|)\xi^{\a-1}
e^{-(a_0-\eta)\xi^{\a}}d\xi]
\] 
with 
\[
\frac{2}{\pi}E_t[\int_0^{+\infty}\bar{H}(\xi,\betamax)\xi^{\a-1}e^{-(a_0-\eta)\xi^{\a}}d\xi].
\]

Finally, it is worth presenting a bound of the same style as (\ref{eq19}), 
entirely depending on  exponential functions:
\begin{equation*}\label{24nuovo}
\begin{split}
K(F(\cdot,t),G_\a) \leq &  \frac{8}{5\pi}M^2N_{2\a}e^{-t(1-2A_{4})} +  \frac{64}{25\pi}M^4N_{4\a}e^{-t(1-2A_{6})}\\
+ &  \frac{e^{-t}}{2}\sup_{x \in \RE}|F_0(x)+F_0(-x-0)-1| \\
&+ \Big(\frac{2}{\pi}\|v_0^*\|(N_\a+2N_{2\a}\|v_0^*\|)+\frac{  \bfc||h_{\a}||}{\tilde{d}}
\Big)e^{-t(1-q\sigma\a/2-2A_q)}\\
+ & e^{-\r t}\frac{\bfc ||h_{\a}||}{\tilde{d}}+  \frac{2}{\pi}\int_0^{+\infty}\bar{H}(\xi,e^{-t\sigma})\xi^{\a-1}e^{-(a_0-\eta)\xi^{\a}}d\xi.\\
\end{split}
\end{equation*}

Notice that the above two bounds go to zero as $t \rightarrow +\infty$. 
Indeed, the latter  goes to zero since, on the one hand,  $\lim_{t \rightarrow +\infty}
\int_0^{+\infty}\bar{H}(\xi,e^{-t\sigma})\xi^{\a-1}e^{-(a_0-\eta)\xi^{\a}}d\xi=0$ and, on the other hand, 
 $\sigma$ and $q$ can be chosen in  such a way that $1-q\sigma\a/2-2A_q$ turns out to be strictly positive. 
Exponential bounds can be given under the usual 
condition on the behavior of $v_0$ near the origin.
\begin{theorem}\label{thm5} 
If, besides the assumptions considered in {\rm Theorem \ref{thm4}}, $\vzs$ is such that $|\vzs(\xi)|=O(|\xi|^{\d})$ as $\xi \rightarrow 0$ for some $\d
>0$, and $d$ is chosen to assure that $|\xi|\leq \tilde{d}=(3d/8m)^{1/\a}$ entails $|\vzs(\xi)|\leq \r |\xi|^{\d}$, 
then
\[
\begin{split}
K(F(\cdot,t),G_\a) \leq & \frac{8}{5\pi}M^2N_{2\a}e^{-t(1-2A_{4})} + \frac{64}{25\pi}M^4N_{4\a}e^{-t(1-2A_{6})}\\
+&  \frac{2}{\pi}\r N_{\a+\d} e^{-t(1-2A_{2+2\d/\a})}+ 2\r^2  N_{2\a+2\d}e^{-t(1-2A_{2+4\d/\a})}\\
+& \frac{\bfc ||h_{\a}||}{\tilde{d}}E_t(\betamax) + \frac{e^{-t}}{2}\sup_{x \in \RE}|F_0(x)+F_0(-x-0)-1|.
\end{split}
\]
\end{theorem}

%%QUI: nel teorema nella forumla seconda riga A_{4+4\delta/\alpha} credo sia A_{2+4\delta/\alpha}
% l'ho cambiato, ricontrollare. 

In view of (\ref{eq18}), the thesis of Theorem \ref{thm5} can be formulated as: {\it There are positive constants $a_3$ and
$b$ such that $K(F(\cdot,t),G_\a) \leq a_3e^{-bt}$ for every $t\geq 0$.}

\subsection{Convergence in Kolmogorov's metric (Conditions expressed on the initial probability distribution $F_0$).}\label{Subsec2.3}

A characteristic feature of the results presented until now is that all the assumptions adopted to obtain bounds
--- in particular, extra-conditions to achieve exponentially fast convergence 
--- are formulated in terms of conditions on the 
initial characteristic function. In general, with respect to actual choice of initial data, it is easier and more natural
to assign conditions on $F_0$ than on  $\phi_0$. Apropos of this remark, see the role played by Lemma 6.1 in 
\cite{PulvirentiToscani2004} and Section \ref{Subsec2.4} below. 
With reference to the classical case of independent and identically distributed summands, see, for example, 
\cite{Cramer1962,Cramer1963}, \cite{Hall1981}. Accordingly, the main objective of the rest of the section is to determine bounds for $K(F(\cdot,t),G_\a)$, 
expressed in terms of quantities whose computation is generally easier than the computation 
of characteristic functions,
once either $F_0$ or some approximate form of $F_0$ has been assigned. To pave the way for presentation, let us complement previous
notation given, in particular, in Subsection \ref{Subec_2.1}:
\[
\begin{split}
h^*(x):= & x^{\a}S^*(x)=x^{\a}\{1-F_0^*(x)\}-c_0^*=x^\a F_0^*(-x)-c_0^*  \qquad (x>0) \\
b_1^*(x):= & 2x\int_D^{+\infty}\sin(xu)S^*(u) du \\
\end{split}
\]
where $D$ is some strictly positive number and the integral has to be meant as improper Riemann integral. Moreover,
\[
\begin{split}
B_1:= & 2k_1N_2+8k_1k_2N_{2+\a}, \qquad B_2:= 8k_1^2N_4,  \qquad  B_3:= 4 k_2N_{1+\a}+ 2 N_1 \\
B_4:= & 4 k_2N_{2+\a}+ 2 N_2, \qquad B_5:=\frac{4}{5}M^2N_{2\a}, \qquad B_6:=\frac{32}{25}M^4 N_{4\a}  \\
\end{split}
\]
with
\[
\begin{split}
k_1:= &\int_0^Dx |S^*(x)|dx,   \qquad k_2:= \sup_{x >0}\frac{|b_1^*(x)|}{ x^{\a}}\leq \max\{||\vzs||+2k_1, 2\int_D^{+\infty}|S^*(x)|dx \}
\end{split}
\]
and
\[
\begin{split}
& H_1^*(q):=\int_{0}^1y^{1-\a}|h^*(y/q)|dy, \qquad H_2^*(q):=\int_{1}^{+\infty} y^{-\a}|h^*(y/q)|dy \\
& k_3:=\sup_{q \in (0,1)}H^*_1(q), \qquad k_4:=\sup_{q \in (0,1)}H^*_2(q). \\
\end{split}
\]

\begin{theorem}\label{thm6}
If $F_0$ belongs to the domain of normal attraction of  $G_\a$ with $\a=2/(1+p)$ in $[1,2)$, 
and $\int_\RE|S^*(x)|dx<+\infty$ if $\a=1$, then 
\[
\begin{split}
K(F(\cdot,t),G_\a) \leq & \frac{2}{\pi}E_t\big[\sum_{j=1}^{\nu_t}|\beta_{j,t}|^{\a}
\{ B_3H^*_1(|\beta_{j,t}|)+B_4 H^*_2(|\beta_{j,t}|) \}  \big]\\
+& \frac{{\bfc}||h_{\a}||}{\tilde{d}}E_t(\betamax)+ \frac{2}{\pi}
\Big \{ B_1 e^{-t(1-2A_{4/\a})} + B_2 e^{-t(1-2A_{(8-2\a)/\a})} \\
& B_5 e^{-t(1-2A_{4})} +  B_6e^{-t(1-2A_{6})} \Big \}
+  \frac{e^{-t}}{2}\sup_{x \in \RE}|F_0(x)+F_0(-x-0)-1|. \\
\end{split}
\]
\end{theorem}

Then, setting $\bar H_i^*(x):=\sup_{y \leq x} H_i^*(y)$ for $i=1,2$, and recalling 
(\ref{eq18}), we obtain a bound completely expressed in terms of  exponential functions, that is
\[
\begin{split}
K(F(\cdot,t),G_\a) \leq & \frac{2}{\pi} \Big \{ B_1 e^{-t(1-2A_{4/\a})}
+ B_2 e^{-t(1-2A_{(8-2\a)/\a})}     \\
&
+ \Big (k_3B_3+k_4B_4 +\frac{\pi}{2} \frac{{\bfc}||h_{\a}||}{\tilde{d}} \Big ) e^{-t(1-q\s\a/2-2A_q)}  \\
& +B_5 e^{-t(1-2A_{4})} +  B_6e^{-t(1-2A_{6})} +\frac{\pi}{2} \frac{{\bfc}||h_{\a}||}{\tilde{d}}
e^{-\s t} \\
&+ B_3\bar H_1^*(e^{-\s t}) +B_4\bar H_2^*(e^{-\s t})\Big \} 
+  \frac{e^{-t}}{2}\sup_{x \in \RE}|F_0(x)+F_0(-x-0)-1|. \\
\end{split}
\]

In order to obtain exponential bounds, we reinforce the assumptions made in Theorem~\ref{thm6}, in the sense that
\begin{equation}\label{eq20}
|h^*(x)| \leq \frac{\rho'}{|x|^\delta}  \qquad \text{for some positive constant $\rho'$ and $\delta$ in $(0,2-\a)$}.
\end{equation}

\begin{theorem}\label{thm7} Besides the assumptions made in {\rm Theorem \ref{thm6}}, suppose {\rm(\ref{eq20})}
holds true. Then,
\[
\begin{split}
K(F(\cdot,t),G_\a) \leq & \frac{2}{\pi} \Big \{ B_1 e^{-t(1-2A_{4/\a})}
+ B_2 e^{-t(1-2A_{(8-2\a)/\a})}     +B_5 e^{-t(1-2A_{4})}  \\
&+  B_6e^{-t(1-2A_{6})}+  \big(\frac{\rho'B_3}{2-\a-\delta}+\frac{\rho'B_4}{\a+\delta-1}\big) e^{-t(1-2A_{2+2\delta/\a})} \Big\} \\
&+  \frac{{\bfc}||h_{\a}||}{\tilde{d}}(e^{-\s t}+e^{-t(1-q\s\a/2-2A_q)} )
+  \frac{e^{-t}}{2}\sup_{x \in \RE}|F_0(x)+F_0(-x-0)-1| \\
\end{split}
\]
which is tantamount to saying that there are positive constants $a_4$ and $b_4$ such that 
$K(F(\cdot,t),G_\a) \leq a_4 e^{-b_4t}$ holds for every $t \geq 0$.
\end{theorem}

It remains to consider the case with $\a$ in  $(0,1)$. In point of fact, 
the next theorem is valid for any $\a$ in $(0,2)$, but it requires further notation.
Firstly, $S^{*}$ is assumed to be {\it monotonic} on $[D,+\infty)$. Then, one sets
\[
\begin{split}
&b_2^*(x):=-2\int_D^{+\infty}(1-\cos(xy))dS^*(y);\\
&H_3^*(q):=\int_{1}^{+\infty} y^{-(1+\a)}|h^*(y/q)|dy, 
\quad \bar H_3^*(q):=\sup_{y\leq q}H_3^*(y), \quad k_5:=\sup_{q \in (0,1)}H_3^*(q);\\
&\bar B_1 := 2\bar k_1 N_2+8\bar k_1 \bar k_2 N_{2+\a} + |S^*(D)|D^2N_2+2\bar k_2|S^*(D)|D^2N_{2+\a}, \\ 
&\bar B_2 := 8 \bar k_1^2 N_4,  \qquad  \bar B_3:= 2z_0 +4z_\a \bar k_2 \\
\end{split}
\]
with
\[
\begin{split}
& \bar k_1:= k_1+ \frac{D^2|S^*(D)|}{2}, \\
& \bar k_2:=\sup_{x >0}\frac{|b_2^*(x)|}{ x^{\a}} \leq  k_2 +2 D |S^*(D)|\max\Big(\frac{D^2}{2},2\Big), \\
& z_r:= \max\Big \{\int_0^{+\infty} \Big |\frac{d}{dx}n_r(x)\Big |dx, \frac{1}{2} \int_0^{+\infty} x^2 \Big|\frac{d}{dx}
n_r(x)\Big|dx  \Big \}\\
\end{split}
\]
where
\[
n_r(x):=e^{-(a_0-\eta)x^\a}x^r \qquad x >0.
\]

\begin{theorem}\label{thm8}
Let $\a$ belong to $(0,2)$ and let $S^*$ be monotonic on $[D,+\infty)$. Then,
\[
\begin{split}
K(F(\cdot,t),G_\a) \leq & \frac{2}{\pi} \bar B_3 E_t\big[\sum_{j=1}^{\nu_t}|\beta_{j,t}|^{\a}
 \{H^*_1(|\beta_{j,t}|)+H^*_3(|\beta_{j,t}|)\}  \big]\\
+& \frac{{\bfc}||h_{\a}||}{\tilde{d}}E_t(\betamax)+ \frac{2}{\pi}
\Big \{ \bar B_1 e^{-t(1-2A_{4/\a})} + \bar B_2 e^{-t(1-2A_{(8-2\a)/\a})} \\
& B_5 e^{-t(1-2A_{4})} +  B_6e^{-t(1-2A_{6})} \Big \}
+  \frac{e^{-t}}{2}\sup_{x \in \RE}|F_0(x)+F_0(-x-0)-1|. \\
\end{split}
\]
\end{theorem}

As done elsewhere in this section, it should be noted that the inequality
\[
\begin{split}
\frac{{\bfc}||h_{\a}||}{\tilde{d}}E_t(\betamax)& + \frac{2}{\pi} \bar    B_3E_t\big[\sum_{j=1}^{\nu_t}|\beta_{j,t}|^{\a}
 \{H^*_1(|\beta_{j,t}|)+H^*_3(|\beta_{j,t}|)\}  \big] \\
&\leq  \left ( \frac{{\bfc}||h_{\a}||}{\tilde{d}} +\frac{2}{\pi} \bar   B_3(k_3+k_5) \right ) e^{-t(1-q\s \a/2-2A_q)} \\
&+ \frac{{\bfc}||h_{\a}||}{\tilde{d}} e^{-\s t} +\frac{2}{\pi} \bar B_3 \{\bar H_1^*( e^{-\s t})+ 
\bar H_3^*( e^{-\s t}) \}\\
\end{split}
 \]
is useful to yield a bound for $K(F(\cdot,t),G_\a) $ depending only on exponential functions, 
while an exponential bound can be derived from the next theorem.

\begin{theorem}\label{thm9} Besides the assumptions made in {\rm Theorem \ref{thm8}}, suppose {\rm(\ref{eq20})}
holds true. Then, 
\[
\begin{split}
K(F(\cdot,t),G_\a) \leq & \frac{2}{\pi} \Big \{ \bar B_1 e^{-t(1-2A_{4/\a})}
+ \bar B_2 e^{-t(1-2A_{(8-2\a)/\a})}     +B_5 e^{-t(1-2A_{4})}  \\
&+  B_6e^{-t(1-2A_{6})}+  \big(\frac{\rho'\bar B_3}{2-\a-\delta}+\frac{\rho'\bar B_3}{\a+\delta}\big) e^{-t(1-2A_{2+2\delta/\a})} \Big\} \\
&+  \frac{{\bfc}||h_{\a}||}{\tilde{d}}(e^{-\s t}+e^{-t(1-q\s\a/2-2A_q)} )
+  \frac{e^{-t}}{2}\sup_{x \in \RE}|F_0(x)+F_0(-x-0)-1|. \\
\end{split}
\]
\end{theorem}

\subsection{Brief comparative study of extra--condition on $\phi_0$ and on $F_0$}\label{Subsec2.4}
In view of the greater expressiveness of assumptions given for $F_0$, if compared to 
conditions on 
$\phi_0$, already stressed at the beginning of Subsection \ref{Subsec2.3}, we conclude the section with a brief comparative analysis. This analysis deals, on the one hand, with the two kinds of conditions actually used 
in the present paper and, on the other hand, with our conditions on  $F_0$ 
compared with those introduced in \cite{PulvirentiToscani2004}.

Recall that in Subsections 2.1 and 2.2 we have used an extra--condition which, in the symmetric case, reduces to
\begin{equation}\label{eq21A}
|\vzs(\xi)|=O(|\xi|^\delta) \qquad \text{as $\xi\to 0$, \,\,  for some $\d>0$}
\end{equation}
while, in Subsection 2.3, we have stated a few results under the extra--condition 
\begin{equation}\label{eq22A}
\Big |(1-F_0^*(x))-\frac{c_0^*}{x^\a} \Big  | \leq \frac{\rho'}{x^{\a+\d}} \qquad (x >0)
\end{equation}
for some $\d$ in $(0,2-\a)$ when $\a$ belongs to $[1,2)$, and for some $\d$ in 
$(0,2-\a)$ when $\a$ belongs to $(0,1)$ provided that $S^*(x)=(1-F_0^*(x))-c_0^*x^{-\a}$
is monotonic for $x>D\geq0$. 

As to the former point under discussion, notice that for $\a$ in $[1,2)$ one can resort to easy inequalities, to be 
explained and used in  the proof of Proposition \ref{prop5}, to obtain 
\[
|\vzs(\xi)| \leq \frac{|b_1^*(\xi)|}{|\xi|^\a} +2 k_1 |\xi|^{2-\a}
\]
where, in view of (\ref{eq22A}), $|b_1^*(\xi)|=O(|\xi|^{\a+\d})$. An analogous conclusion holds true
when $0<\a<1$ with $b_2^*$ and $\bar k_1$ in the place of $b_1^*$ and $k_1$, respectively. 
See formal developments in  the proof of Proposition \ref{prop6}. Hence: {\it If $\d$ belongs to $(0,2-\a)$
with $0<\a<2$, and $S^*$ is monotonic on $(D,+\infty)$ for some $D \geq 0$ when $0<\a<1$, then
{\rm(\ref{eq22A})} entails {\rm(\ref{eq21A})}.}

Moving on to the latter kind of comparisons, it should be recalled that \cite{PulvirentiToscani2004}, 
in order that  initial data can satisfy (\ref{eq22A}), assume
\begin{equation}\label{eq23A}
m_{\a+\d}:=\int_\RE |x|^{\a+\d} |f_0(x)-\galpha(x)|dx <+\infty \qquad \text{ for some $\d>0$  }.
\end{equation}
In Section 4 of \cite{GoudonJuncaToscani2002}
it is proved that (\ref{eq23A}) {\it entails}  (\ref{eq21A}) and now we prove that 
(\ref{eq23A}) {\it yields}  (\ref{eq22A}) {\it when $\d \leq \a$}. Indeed, from the Markov inequality,
\[
|F^*(x)-G_\a(x)| \leq \frac{m_{\a+\d}}{2x^{\a+\d}}.
\]
This, combined with a well--known asymptotic expression for $G_\a$ (see, for example, Sections 2.4 
and 2.5 of \cite{Zolotarev1986}), gives
\[
|F^*(x)-\frac{c_0^*}{x^\a}| \leq \frac{m_{\a+\d}}{2x^{\a+\d}} +O(\frac{1}{x^{2\a}}) 
\qquad (x \to +\infty).
\]
Then, (\ref{eq22A}) {\it follows form} (\ref{eq23A}) {\it when $\d \leq \a$}. This last restriction
is consistent with Theorem 6.2 in \cite{PulvirentiToscani2004}, 
mentioned in $(F_5)$, and with the first part of Theorem \ref{thm3}. Moreover, it should be noted
that classical asymptotic formulae for $\galpha$ (see, e.g., \cite{IbragimovLinnik1971}) 
can be applied to exhibit simple examples of initial data which meet  (\ref{eq22A})
but do not meet  (\ref{eq23A}). In other words, the criterion evoked by \cite{PulvirentiToscani2004}
-- to get (\ref{eq21A}) together with exponential bounds for $\chi_{\a+\d}$ with $\d \leq \a$ --
 could be 
usefully replaced by the weaker condition (\ref{eq22A}), as we have done for convergence with respect to the Kolmogorov metric.

\section{Limit theorems for weighted sums of independent random numbers}\label{s3}

As mentioned in the introductory paragraph of Section \ref{s2} ---
 see, in particular, explanation for
(\ref{eq14}), (\ref{eq15_1}) and (\ref{eq15_2}) --- the present section focuses on the study of the convergence in distribution of weighted 
sums of independent random variables. This study, besides the interest it could hold in itself, is essential for proving
the theorems already stated in Section \ref{s2}. In point of fact, the main steps of the arguments used to prove these theorems are set
out in the propositions we get ready to enunciate and prove in the present section. Specific indications of how they are used will be given in the Appendix.

For the present, it should be recalled that we are interested in convergence in distribution of sums %(\ref{eq14}),
\begin{equation}\label{eq21}
S_n:=\sum_{j=1}^n q^{(n)}_jX_j
\end{equation}
with $X_1, X_2, \dots$ {\it independent and identically distributed real--valued random variables with common distribution function
$F_0$}. Moreover,  the numbers $ q^{(n)}_j$ {\it are assumed to satisfy} (\ref{eq15_1}) - (\ref{eq15_2}), and $F_0$ is supposed 
to be a {\it symmetric element
of the domain of normal attraction of} (\ref{eq7}). Then according to ($F_3$) and ($F_4$), there is $c_0\geq 0$ satisfying
\begin{equation}\label{eq21_bis}
a_0=2c_0\int_0^{+\infty}\frac{\sin(x)}{x^{\a}}dx
\end{equation}
for which
\[
\lim_{x\rightarrow -\infty}|x|^{\a}F_0(x)=\lim_{x\rightarrow +\infty}x^{\a}\{1-F_0(x)\}=c_0
\]
and
\[
1-\phi_0(\xi)=(a_0+v_0(\xi))|\xi|^{\a}  \qquad (\xi \in \RE)
\]
where $v_0$ is a bounded real-valued function satisfying   $|v_0(\xi)|=o(1)$ as $\xi \rightarrow 0$. 
See (\ref{eq12bis}) - (\ref{eq12ter}).

The above conditions, printed in italic type, are assumed to be in force throughout the present section, and will be not
repeated in the following statements. It is worth recalling that these statements are inspired by previous work published in \cite{Cramer1962,Cramer1963} and \cite{Hall1981}. 
Accordingly, the present line of reasoning is based on certain inequalities contained in
the following lemma where, as in the rest of the section, 
for the sake of typographic convenience, $q_j$ is used instead of $q_j^{(n)}$.

\begin{lemma}\label{lemma1}
Let $\tilde{\phi}_n$ be the characteristic function of {\rm (\ref{eq21})}. Then,
\begin{equation}\label{eq22}
\begin{split}
|\tilde{\phi}_n(\xi) &- \trgal(\xi)| \leq  e^{-(a_0-\eta)|\xi|^{\a}} |\xi|^{\a} \Big
\{ \sum_{j=1}^{n}q_j^{\a}|v_0(\xi q_j)|
(1+2|\xi|^{\a}|v_0(\xi q_j)|)\\
+ & |\xi|^{\a}M^2\sum_{j=1}^{n}q_j^{\a} \Big(\frac{4}{5}q_j^{\a}
+\frac{32}{25}M^2|\xi|^{2\a}q_j^{2\a} \Big)\Big\}\J \{|\xi|\leq D_n\}\\
+&  2\J \{|\xi|> D_n\} \Big(\frac{|\xi|}{d_1} \Big)^s \Big[\frac{q_{(n)}^s}{d^{s/\a}}\J \{c=0\} 
+ \frac{q_{(n)}^{s}}{d^{s/\a}}\J\{
q_{(n)}> d^{1/c\a}, \; 0<c<1\} \\& 
+ q_{(n)}^{s(1-c)}\J\{q_{(n)}\leq  d^{1/c\a}, 0<c<1\}\Big]\\
\end{split}
\end{equation}
holds for any $\xi$ in $\RE$, $s>0$, $c$ in $[0,1)$, $d$, $d_1$, $k^*$ and 
$M$ being the same as in {\rm Theorem \ref{thm2}} with $v_0$ in the place of $v_0^*$,
$q_{(n)}=max\{q_1, \dots q_n\}$ and $D_n=D_n(c,d):=(\frac{3}{8M}(d\wedge q_{(n)}^{c\a}))^{1/\a}q_{(n)}^{-1}\J\{0<c<1\}
+ (\frac{3}{8M}d)^{1/\a}q_{(n)}^{-1}\J \{c=0\}$.  Moreover, for $s=\a$, $c$ in $(0,1)$ and $\xi$ in $\RE$,
\begin{equation}\label{eq23}
\begin{split}
|\tilde{\phi}_n(\xi)- \trgal(\xi)| \leq & |\xi|^{\a} \Big [
e^{-(a_0-\eta)   |\xi|^{\a}}  \Big (  k^*\J \Big\{ q_{(n)}> d, \; |\xi|\leq 
\frac{d_1d^{1/\a}}{q_{(n)}} \Big\}\\
+ & \bar{\sigma}(\xi) q_{(n)}^\a \J \Big \{q_{(n)}\leq  d^{1/c\a}, |\xi|\leq d_1 q_{(n)}^{c-1} \Big \}\Big )\\
& + 
\frac{2}{d_1^{\a}}\Big (\frac{q_{(n)}^\a}{d} \J \{
q_{(n)}> d^{1/c\a}\}+ q_{(n)}^{\a(c-1)} \Big) \Big ]
 \end{split}
\end{equation}
with
\[
\begin{split}
\bar{\sigma}(\xi)= \sum_{j=1}^{n}q_j^{\a}|v_0(\xi q_j)|+|\xi|^{\a}\Big(\frac{4}{5}M^2  \sum_{j=1}^{n}q_j^{2\a}+2 \sum_{j=1}^{n}q_j^{\a}|v_0(\xi q_j)|^2+ \frac{32}{25}|\xi|^{3\a}M^4 \sum_{j=1}^{n}q_j^{3\a}\Big).
\end{split}
\]
\end{lemma}

{\it Proof.} According to previous notation, set
$\|v_0\|:=\sup_{\{x>0\}}|v_0(x)|$ and $\bar v_0(\xi):=\sup_{\{0<x\leq \xi\}}|v_0(x)|$.
Now, in view of $(F_4)$,
\[
|1-\phi_0(\xi q_j)| = |a_0+v_0(\xi q_j)||\xi q_j|^\a \leq M|\xi|^\a q_j^\a 
\]
and the last term turns out to be bounded from above by $3d/8 \leq 3/8$
when $|\xi|q_{(n)} \leq (3d/8M)^{1/\a}$.
 Since 
$\log(1+z)=z+(4/5) \th_z |z|^2$ for $|z| \leq 3/8$ and some $\th_z$ satisfying $|\th_z|\leq 1$ (see, for example, 
Lemma 3 in Section 9.1 of \cite{ChowTeicher1997}), then $|\xi|q_{(n)} \leq  (3d/8M)^{1/\a}$
yields 
\[
\begin{split}
\tilde \phi_n(\xi)&=\exp\{\sum_{j=1}^n \log(\phi_0(\xi q_j))\} 
=\exp\{ \sum_{j=1}^n \log(1-(1-\phi_0(\xi q_j))) \} \\
&= \exp \{ -\sum_{j=1}^n (1-\phi_0(\xi q_j)) + \sum_{j=1}^n r(1-\phi_0(\xi q_j))\}
\end{split}
\]
with $r(x):=(4\th_x/5)|x|^2$. 
Moreover, if  $|\xi|\leq (3d/8M)^{1/\a}/q_{(n)}$ and $0<d<1$,
\[
|r(1-\phi_0(\xi q_j))| \leq \frac{4}{5} M^2|\xi|^{2\a} q_j^{2\a}  \qquad (j=1,\dots,n)
\]
and, via $(F_4)$,
\begin{equation}\label{3322}
\begin{split}
\tilde \phi_n(\xi)& =\exp \Big ( -\sum_{j=1}^n \{a_0+v_0(\xi q_j) \}|\xi q_j|^\a 
-\sum_{j=1}^n r(1-\phi_0(\xi q_j ))  \Big ) \\
& =\exp(-a_0|\xi|^\a) \exp (-B_n(\xi)+R_{1,n}(\xi))\\
\end{split}
\end{equation}
with
\[
B_n(\xi)=|\xi|^\a \sum_{j=1}^n q_j^\a v_0(\xi q_j)
\]
and
\[
|R_{1,n}(\xi)|=|\sum_{j=1}^n r(1-\phi_0(\xi q_j ))|\leq \frac{4}{5}M^2|\xi|^{2\a} \sum_{j=1}^n q_j^{2\a}.
\]
Writing
\[
\begin{split}
\exp(-B_n(\xi) + R_{1,n}(\xi))& = 1- B_n(\xi) + R_{1,n}(\xi) \\
&+ (R_{1,n}(\xi)- B_n(\xi)  )^2 \sum_{l \geq 0} \frac{(R_{1,n}(\xi)- B_n(\xi))^l}{l!}\frac{l!}{(l+2)!} \\
&=1-B_n(\xi)+ R_{1,n}(\xi) +R_{2,n}(\xi),
\end{split}
\]
with
\begin{equation}\label{eq24}
\begin{split}
|R_{2,n}(\xi)| &=(R_{1,n}(\xi)- B_n(\xi)  )^2 \Big |\sum_{l \geq 0} \frac{(R_{1,n}(\xi)- B_n(\xi))^l}{l!}\frac{l!}{(l+2)!}\Big| 
\\ &\leq  2 \{B_n(\xi)^2+R_{1,n}(\xi)^2\}\exp(|B_n(\xi)|+|R_{1,n}(\xi)|), \\
\end{split}
\end{equation}
equalities (\ref{3322}) give
\begin{equation}\label{eq25}
\tilde \phi_n(\xi)=\exp(-a_0|\xi|^\a)\{1-B_n(\xi)+ R_{1,n}(\xi) +R_{2,n}(\xi)\}.
\end{equation}
As to $R_{2,n}(\xi)$, for $|\xi|^\a\leq (3d/8M)q_{(n)}^{-\a}$ and any sufficiently small $d$, one gets
\[
\begin{split}
|B_n(\xi)|+|R_{1,n}(\xi)|& \leq
|\xi|^\a \sum_{j=1}^n \bar v_0(\xi q_{(n)}) q_j^\a + \frac{4}{5}M^{2} |\xi|^{2\a} q_{(n)}^\a
\sum_{j=1}^n q_j^\a \\
&\leq |\xi|^\a\{\bar v_0(\xi q_{(n)}) +  \frac{4}{5}M^{2}|\xi|^\a q_{(n)}^\a \} \leq \eta |\xi|^\a \\
\end{split}
\]
by the definition of $d$ given immediately before the beginning of Subsection \ref{Subsec2.1}. This entails
\[
\exp(|B_n(\xi)|+|R_{1,n}(\xi)|) \leq e^{\eta |\xi|^\a}
\]
for any $\eta$ in $(0,a_0)$ and $|\xi| \leq (3d/8M)^{1/\a}  q_{(n)}^{-1}$. 
Next, an application of Jensen's inequality yields 
\[
|B_n(\xi)|^2+|R_{1,n}(\xi)|^2  \leq 
|\xi|^{2\a} \sum_{j=1}^n  q_j^\a v_0(\xi q_j)^2 + \frac{16}{25} M^4 |\xi|^{4\a} \sum_{j=1}^n
 q_j^{3\a}
\]
which, in turn, combined with (\ref{eq24}), gives
\[
|R_{2,n}(\xi)|\leq \Big \{2 |\xi|^{2\a} \sum_{j=1}^n  q_j^\a v_0(\xi q_j)^2 + \frac{32}{25} M^4 |\xi|^{4\a} \sum_{j=1}^n
 q_j^{3\a}  \Big \} e^{\eta|\xi|^\a}.
\]
Now, from (\ref{eq25}) with $|\xi| \leq D_n$,
\[
\begin{split}
|\tilde \phi_n(\xi)-&\exp(-a_0|\xi|^\a)| \leq e^{-a_0|\xi|^\a} |\xi|^\a
\Big \{ \sum_{j=1}^n |v_0(\xi q_{(n)})| q_j^\a \\ 
& +  \frac{4}{5}M^2|\xi|^{\a} \sum_{j=1}^n q_j^{2\a} +
 \Big (2 |\xi|^{\a} \sum_{j=1}^n  q_j^\a v_0(\xi q_j)^2 + \frac{32}{25} M^4 |\xi|^{3\a} \sum_{j=1}^n
 q_j^{3\a}  \Big ) e^{\eta |\xi|^\a}  \Big \} \\
 & \leq e^{-(a_0-\eta)|\xi|^\a} |\xi|^\a 
 \Big \{ \sum_{j=1}^n  q_j^\a |v_0(\xi q_j)|\Big(1+2|\xi|^\a |v_0(\xi q_j)|\Big) \\
 & \qquad \qquad +|\xi|^\a M^2 \sum_{j=1}^n  q_j^{2\a}\Big(\frac{4}{5}+\frac{32}{25} M^2|\xi|^{2\a} q_j^\a \Big) \Big \}.\\
 \end{split}
\]
At this stage it remains to consider $|\xi|>D_n$. In this case, one gets
\[
\frac{|\xi|^s}{d_1^s} \left \{ \frac{q_{(n)}^s}{d^{s/\a}}\J(c=0) + \frac{q_{(n)}^s}{d^{s/\a}} \J
(q_{(n)}> d^{1/c\a}, 0<c<1) +q_{(n)}^{s(1-c)} \J(q_{(n)} \leq d^{1/c\a}, 0<c<1 )   \right   \} \geq 1
\]
and, to complete the proof for (\ref{eq22}), it is enough to take account of the obvious inequality
$|\tilde \phi_n(\xi)-\exp(-a_0|\xi|^\a)|\leq 2$. 

Now, as far as (\ref{eq23}) is concerned, take $s=\a$ and $c$ in $(0,1)$. Then, (\ref{eq22})
becomes
\begin{equation}\label{eq26}
\begin{split}
|\tilde \phi_n(\xi)&-\exp(-a_0|\xi|^\a) | 
\leq e^{-(a_0-\eta)|\xi|^\a}|\xi|^\a \bar \sigma(\xi) \J\{|\xi|\leq D_n \} \\ &+ 
2\frac{|\xi|^\a}{d_1^\a}\Big\{ \frac{q_{(n)}^\a}{d} \J(q_{(n)} >d^{1/c\a} ) 
+q_{(n)}^{\a(1-c)} \J(q_{(n)} \leq d^{1/c\a} )  \Big  \}  \J\{|\xi|> D_n \}. \\
\end{split}
\end{equation}
Now, for $ q_{(n)} >d$ and $|\xi| \leq D_n (\leq d_1d^{1/\a}q_{(n)}^{-1}) $, 
\[
\bar \s(\xi) \leq 
\bar v_0(d_1 d^{1/\a})(1+2d_1^\a d^{1-\a}
\bar v_0(d_1 d^{1/\a})) + (4/5) M^2 d_1^\a d +(32/25) M^4 d_1^{3\a} d^{3-\a}=k^*
\]
and (\ref{eq23}) follows from (\ref{eq26}) with $\bar \s(\xi)$ replaced by $k^*$ on 
$\{ q_{(n)} >d , |\xi| \leq D_n  \}$. $\qquad \fine$

%One can combine (\ref{eq15_1}), (\ref{eq15_2}) 
% and $(F_1)-(F_4)$ to conclude that 
% {\it the upper bounds stated in} (\ref{eq22}) {\it and}
%  (\ref{eq23}), {\it respectively, go to zero  as} $ n \to +\infty$.
  
  Lemma \ref{lemma1} can be used to obtain bounds for the 
  $\chi_\a$--distance between $G_\a$  and the probability distribution function $F_n$
  of $S_n$.

\begin{proposition}\label{prop1}
The $\chi_\a$--distance between $F_n$ and $G_\a$ satisfies
\[
\begin{split}
\chi_\a(F_n,G_\a) & \leq k^* \J(q_{(n)} > d) + \sum_{j=1}^n q_j^\a \bar v_0\big(d_1q_j q^{c-1}_{(n)}\big) \{1+2M_1 \bar v_0\big(d_1q_j q^{c-1}_{(n)}\big)\} 
 \\
&+  q^{\a}_{(n)} \Big \{\frac{4}{5} M_1 M^2 +\frac{32}{25} M_3 M^4 
 q^{\a}_{(n)} \Big \} 
  + \frac{2}{d_1^\a} \Big \{ \frac{ q^{\a}_{(n)}}{d} \J(  q_{(n)} > d^{1/c\a}) +  q^{\a(1-c)}_{(n)} \Big \}
\\
\end{split}
\]
for any $c$ in $(0,1)$, with $M_r:=\max_{x \geq 0} e^{-(a_0-\eta)x^\a}x^{r\a}$ {\rm(}$r$ being any positive number{\rm)}.
\end{proposition}
  
{\it Proof.} Consider (\ref{eq23}) and observe that 
\[
\bar \s(\xi) \leq \sum_{j=1}^n q_j^\a \bar v_0(d_1 q_j q_{(n)}^{c-1})
\Big(1+2M_1\bar v_0(d_1q_jq_{(n)}^{c-1})\Big) 
+\frac{4}{5} M^2M_1 q_{(n)}^\a+\frac{32}{25} M^4 M_3 q_{(n)}^{2\a}
\]
holds true on the   set $\{q_{(n)} \leq d, |\xi| \leq D_n  \}$ since 
$D_n \leq d_1 q_{(n)}^{c-1}$ on this set. $\quad \fine$  

It is easy to check that the upper bound stated in Proposition \ref{prop1}
is $o(1)$ for $n \to +\infty$. 

Lemma \ref{lemma1} can also be exploited to determine analogous bounds for
$\chi_{\a+\d}$ and $\chi_{2\a}$, under the extra--condition (\ref{eq:13bis}). 
%\begin{equation}\label{eq27}
%|v_0(|\xi|)|=O(|\xi|^\d) \quad \text{as $\xi \to 0$, for some $\d>0$.}
%\end{equation}

\begin{proposition}\label{prop2}  Suppose {\rm(\ref{eq:13bis})} is valid for some $\d>0$ and 
take $d$  in such a way 
that $|\xi|q_{(n)} \leq d_1 d^{1/\a}$ $(=q_{(n)}D_n$ if $c=0)$ entails 
$\bar v_0(\xi q_j) \leq \rho |\xi q_j|^\delta$ for some $\rho>0$. Then,
\[
\begin{split}
\chi_{\a+\d}(F_n,G_\a) &\leq \rho  \sum_{j=1}^n q_j^{\a+\d}    +2\rho^2 M_{1+\frac{\d}{\a}}  \sum_{j=1}^n q_j^{\a+2\d}     \\
& +\frac{4}{5}M^2 M_{1-\frac{\d}{\a}} \sum_{j=1}^n q_j^{2\a} + \frac{32}{25}M^4 M_{3-\frac{\d}{\a}} 
\sum_{j=1}^n q_j^{3\a} + \frac{2q_{(n)}^{\a+\d}}{d_1^{\a+\d}d^{1+\d/\a}} \\
\end{split}
\]
for any $\d \leq \a$, and
\[
%\begin{split}
\chi_{2\a}(F_n,G_\a) %& 
\leq  \rho M_{\frac{\d}{\a}-1}  \sum_{j=1}^n q_j^{\a+\d} 
+2\rho^2 M_{\frac{2\d}{\a}}  \sum_{j=1}^n q_j^{\a+2\d} 
%\\
%\\
%&
+\frac{4M^2}{5} \sum_{j=1}^n q_j^{2\a}
+\frac{32M^4M_{2}}{25}  \sum_{j=1}^n q_j^{3\a}
+ \frac{2q_{(n)}^{2\a}}{d_1^{2\a}d^{2}} 
%\\
%\end{split}
\]
for any $\d$ in $(\a,2\a]$.
\end{proposition}

{\it Proof.}
From (\ref{eq22}) with $c=0$ and $s=\a+\d$,
\[
\begin{split}
|\tilde \phi_n(\xi)-&e^{-a_0|\xi|^\a}| \leq e^{-(a_0-\eta)|\xi|^\a}|\xi|^\a \Big \{ 
  \rho \sum_{j=1}^n q_j^{\a+\d} |\xi|^\delta(1+2\rho q_j^\d|\xi|^{\a+\d}) \\
  &+|\xi|^\a M^2  \sum_{j=1}^n q_j^{2\a}(\frac{4}{5}+\frac{32}{25} M^2q_j^\a|\xi|^{2\a})
  \Big \} \J(|\xi| \leq d_1d^{1/\a}q_{(n)}^{-1}) \\
& + \frac{2q_{(n)}^{\a+\d}}{d_1^{\a+\d}d^{1+\d/\a}} |\xi|^{\a+\d} \J( |\xi| > d_1d^{1/\a}q_{(n)}^{-1}).
\\
\end{split}
\]
Then, if $\d$ belongs to $(0,\a]$, one easily obtains the former of the inequalities to be proved. 
The latter follows similarly from (\ref{eq22}) with $c=0$ and $s=2\a$.
$\quad \fine$

As mentioned at the beginning of Subsection \ref{Subsec2.2}, here we pass from weighted $\chi$--metrics
to Kolmogorov's metric via the classical
{Berry--Esseen inequality}
\[
K(F_n,G_\a) \leq \frac{1}{\pi} \int_{-\tilde{d}/q_{(n)}}^{\tilde{d}/q_{(n)}}
\Big |\frac{\tilde \phi_n(\xi)-\trgal(\xi)}{\xi} \Big |d\xi + \frac{\bfc}{\tilde{d}} \|\galpha \|q_{(n)}
\]
$\bfc$ being the constant which appears in Theorem 3.18 
in \cite{Galambos1995}. 

Take  (\ref{eq22}), with  $c=0$ and $\tilde{d}=(3d/8M)^{1/\a}$, 
and sobstitute it in the  right--hand side of the above Berry--Esseen inequality to obtain

\begin{proposition}\label{prop3}
One has
\begin{equation}\label{eq27_bis}
\begin{split}
K(F_n,G_\a) 
\leq \frac{2}{\pi} \sum_{j=1}^n q_j^\a  &\int_0^{\tilde{d}/q_{(n)}}
e^{-(a_0-\eta)\xi^\a}\xi^{\a-1}H(\xi,q_j) d\xi +
\frac{8}{5\pi}M^2N_{2\a} \sum_{j=1}^n q_j^{2\a} \\
&+ \frac{64}{25\pi} M^4
N_{4\a} \sum_{j=1}^n q_j^{3\a} + \frac{\bfc}{\tilde{d}} \|\galpha \|q_{(n)}\\
\end{split}
\end{equation}
with $H(\xi,q_j):=|v_0(\xi q_j)|(1+2|\xi|^\a|v_0(\xi q_j)|)$ and 
$N_l=\int_0^{+\infty} \exp\{-(a_0-\eta)\xi^\a\}\xi^{l-1}d\xi$. This upper bound is 
$o(1)$ as $n \to +\infty$. 
\end{proposition}

More informative bounds can be obtained under extra--condition (\ref{eq:13bis}).

\begin{proposition}\label{prop4} If {\rm(\ref{eq:13bis})} is valid for some $\d>0$
and $d$ is fixed in such a way 
that $|\xi|q_{(n)} \leq d_1 d^{1/\a}$ $(=q_{(n)}D_n$ if $c=0)$ entails 
$v_0(\xi q_j) \leq \rho |\xi q_j|^\delta$ for some $\rho>0$, then
\[
\begin{split}
K(F_n,G_\a) 
\leq \frac{2}{\pi} & \Big [ \rho N_{\a+\d}
\sum_{j=1}^n q_j^{\a+\d} +2\rho^2  N_{2(\a+\d)} \sum_{j=1}^n q_j^{\a+2\d} +\frac{4}{5} M^2 N_{2\a} \sum_{j=1}^n q_j^{2\a} \\
&
+ \frac{32}{25} M^4
N_{4\a} \sum_{j=1}^n q_j^{3\a} \Big ] + 
\frac{\bfc}{\tilde{d}} \|\galpha \|q_{(n)}=o(1) \quad \text{as $n \to +\infty$.}\\
\end{split}
\]
\end{proposition}

{\it Proof.} Under the present extra--condition, inequality in the previous proposition 
combined with inequality $H(\xi,q_j) \leq \rho|\xi|^\d q_j^\d(1+2\rho q_j^\d|\xi|^\d)$, valid
for every $j$ and $|\xi| \leq \tilde{d}/q_{(n)}$, yields the desired bound. $\qquad \fine$

Now, we proceed to present bounds for $K(F_n,G_\a)$ under restrictions on the initial distribution function, 
rather than on $\phi_0$. Notation is the same as in Subsection \ref{Subsec2.3}
with the proviso that $F_0^*$ is replaced by (symmetric) $F_0$ and, consequently, 
symbols with $*$, like $S^*$, $h^*$, $c_0^*$, etc. must be changed to symbols without
$*$, i.e., $S$, $h$, $c_0$, etc., respectively.

\begin{proposition}\label{prop5}
Let $\a$ be in $[1,2)$ and let the additional 
restriction that $\int_0^{+\infty} |S(x)|dx < +\infty $ if $\a=1$ be valid. Then,
\[
\begin{split}
K(F_n,G_\a) & \leq \frac{2}{\pi} \sum_{j=1}^n  \Big\{
B_1 q_j^{2} + B_2 q_j^{4-\a} + (B_3H_1(q_j)+B_4H_2(q_j))q_j^{\a}\\
& 
+ B_5  q_j^{2\a} +B_6 q_j^{3\a} \Big\} 
+ \frac{\bfc \|\galpha\|}{\tilde{d} } q_{(n)} =o(1) \quad \text{as $n \to + \infty$.} 
\\
\end{split}
\]
In particular, if $h$ is such that $|h(x)|:=x^\a|S(x)| \leq \rho'/x^\d$ for any $x>0$, 
$\d$ in $(0,2-\a)$ and some constant $\rho'>0$, then 
\[
H_1(q) \leq \frac{\rho'q^\delta}{2-\a-\d}, \qquad H_2(q) \leq \frac{\rho'q^\delta}{\a+\d-1}
\]
are valid for any $q$ in $(0,1]$.
\end{proposition}

{\it Proof.} We start from the definitions of $S$ and $\phi_0$ to obtain, via
(\ref{eq21_bis}), 
\[
1-\phi_0(\xi)=a_0|\xi|^\a+2\xi \int_0^{+\infty}S(x) \sin(\xi x)dx
\]
which, in view of $(F_4)$, yields
\[
|\xi|^\a|v_0(\xi q_j)| =\frac{1}{q_j^\a}|b_1(\xi q_j)+R_1(\xi q_j)|
\]
where 
\[
b_1(y):=2y \int_D^{+\infty} \sin(yx) S(x) dx \,\,\,\,\, \text{and}
\,\,\,\,\, R_1(y):=2y \int_0^D \sin(yx) S(x) dx .
\]
For these quantities one can write
\[
|R_1(\xi q_j)| \leq 2\xi^2 q_j^2 \int_0^D x |S(x)| dx =2k_1 \xi^2 q_j^2
\]
with $k_1:=\int_0^D x |S(x)| dx$, and 
\[
k_2:=\sup_{x>0} \frac{|b_1(x)|}{x^\a} \leq \max \{\|v_0\|+2k_1, 2\int _D^{+\infty} |S(x)|dx \}. 
\]
Combination of these inequalities with the definition of $H$ (see Proposition \ref{prop3})
gives us
\[
\begin{split}
|\xi|^{\a-1}&  |H(\xi,q_j)| = |\xi|^{\a-1} |v_0(\xi q_j)|(1+2|\xi|^\a |v_0(\xi q_j)|) 
\\
\leq &\frac{1}{|\xi|q_j^\a} \Big\{ 
|b_1(q_j\xi)|+|R_1(q_j\xi)| +\frac{2}{q_j^\a} \big (|b_1(q_j\xi)|+|R_1(q_j\xi)|   \big )^2\Big \}
\\
\leq &\frac{1}{|\xi|q_j^\a} \Big\{ |b_1(q_j\xi)|+2k_2 |b_1(q_j\xi)||\xi|^\a
+ 2k_1 |\xi q_j|^2 
+ 8 k_1 k_2 q_j^2 |\xi|^{2+\a} +8 k_1^2 q_j^{4-\a}|\xi|^4 \Big \}.
\\
\end{split}
\]
Using this inequality, we obtain

\begin{equation}\label{eq28}
\begin{split}
\frac{2}{\pi} &\sum_{j=1}^n q_j^\a \int_0^{\tilde{d}/q_{(n)}} e^{-(a_0-\eta)\xi^\a} \xi^{\a-1} 
H(\xi,q_j) d\xi \\
& \leq 
\frac{2}{\pi} \sum_{j=1}^n \Big \{  
\int_0^{+\infty} e^{-(a_0-\eta)\xi^\a} \Big ( 
 |b_1(q_j\xi)|\xi^{-1}+2k_2 |b_1(q_j\xi)|\xi^{\a-1} \Big )d\xi \\
 &\qquad \qquad + 2k_1 N_{2} q_j^2 
+ 8 k_1 k_2 q_j^2 N_{2+\a} + 8 k_1^2N_4 q_j^{4-\a} \Big \}.
\\
\end{split}
\end{equation}
It remains to study integrals like $I_r(q)
:=\int_0^{+\infty} |b_1(\xi q)|\xi^{r-1} e^{-(a_0-\eta)\xi^\a}d\xi$
for $r\geq 0$. Following the argument used in \cite{Hall1981}
to prove Lemma 7, one can state the inequality
\begin{equation}\label{eq36BIS}
\begin{split}
I_r(q) &\leq 2q N_{r+2} \int_{\frac{1}{q}}^{+\infty}
|S(x)|dx +2q^2 N_{r+1}\int_0^{\frac{1}{q}}x|S(x)|dx \\
& =2 N_{r+2} q^\a \int_1^{+\infty} |h(y/q)|y^{-\a} dy
+2 N_{r+1} q^\a  \int_0^1 |h(y/q)|y^{1-\a} dy
\\
&=2 N_{r+2} q^\a H_2(q)+ 2 N_{r+1} q^\a  H_1(q)\\
\end{split}
\end{equation}
with $h(x)=x^\a S(x)$. 
To complete the proof of the main part of the proposition it is enough to use
(\ref{eq36BIS}) to obtain a bound for the right-hand side of (\ref{eq28}) and,
then, to replace this bound for the first sum in the right--hand side of (\ref{eq27_bis}).
As to the latter claim, recall that
$H_1(q)=\int_0^1 y^{1-\a}|h(y/q)|dy$, $H_2(q)=\int_1^{+\infty} y^{-\a}|h(y/q)|dy$
and use the additional condition. $\quad \fine$

\begin{proposition}\label{prop6}
Let $\a$ be in $(0,2)$ and let the additional 
hypothesis that $S$ is monotonic on $[D,+\infty)$ be valid for some $D \geq 0$. Then, 
\[
\begin{split}
K(F_n,G_\a) & \leq \frac{2}{\pi} \sum_{j=1}^n  \Big\{
\bar B_1 q_j^{2} + \bar B_2 q_j^{4-\a} + \bar B_3(H_1(q_j)+H_3(q_j))q_j^{\a}\\
& 
+ B_5  q_j^{2\a} +B_6 q_j^{3\a} \Big\} 
+ \frac{\bfc \|\galpha\|}{\tilde{d} } q_{(n)} =o(1) \quad \text{as $n \to + \infty$.} 
\\
\end{split}
\]
Moreover, if $h$ is such that $|h(x)|\leq \rho'/x^\d$ for any $x>0$, 
$\d$ in $(0,2-\a)$ and some constant $\rho'>0$, one gets 
\[
H_1(q) \leq \frac{\rho'q^\delta}{2-\a-\d}, \qquad H_3(q) \leq \frac{\rho'q^\delta}{\a+\d}
\]
for every $q$ in $(0,1]$.
\end{proposition}

{\it Proof.} One starts from Proposition
\ref{prop3} once again, noticing that equality
\[ |t|^{\a}v_0(t)=b_2(t)+R_2(t)
\]
holds with
\[
b_2(t):= -2\int_D^{+\infty}(1-\cos(tx))dS(x)
\,\,\,\, \text{and} \,\,\,\,
R_2(t):= R_1(t)+2 S(D)(\cos(tD)-1).
\]
Observe that
\[
|R_2(\xi q_j)| \leq 2 \bar{k}_1|\xi q_j|^2
\]
with $\bar{k}_1=k_1+D^2|S(D)|/2.$ Moreover,
\[
\bar{k}_2= \sup_{x>0} \frac{|b_2(x)|}{x^{\a}}  \leq k_2+2D|S(D)|\max\Big(\frac{D^2}{2},2\Big).
\]
Then,
\[
\begin{split}
|\xi|^{\a-1}|H(\xi, q_j)| \leq & \frac{1}{|\xi| q_j^{\a}}\Big\{ |b_2(\xi q_j)|+ |R_2(\xi q_j)|+2\frac{(|b_2(\xi q_j)|+ |R_2(\xi q_j)|)^2}{q_j^{\a}}\Big\}\\
\leq & \frac{1}{|\xi| q_j^{\a}}\{ |b_2(\xi q_j)|+2 |\xi|^{\a}\bar{k}_2|b_2(\xi q_j)|+ 2 \bar{k}_1 |\xi q_j|^2 +8\bar{k}_1^2q_j^{4-\a}|\xi|^4+ 8\bar{k}_1\bar{k}_2q_j^2|\xi|^{\a +2} \}.
\end{split}
\]
Hence,
\[
\begin{split}
K(F_n,G_\a) & \leq \frac{2}{\pi} \sum_{j=1}^n  \Big\{\int_0^{\tilde{d}/q_{(n)}} 
\frac{e^{-(a_0-\eta)\xi^\a}}{\xi}[1+2\bar{k}_2 \xi^{\a}]|b_2(\xi q_j)|d\xi \\
 & + (2  \bar{k}_1 N_2+ 8  \bar{k}_1 \bar{k}_2 N_{\a+2})q_j^2+ 8 \bar{k}_1^2N_4q_j^{4-\a} +
\frac{4}{5}M^2 N_{2\a }q_j^{2\a}+ \frac{32}{25}M^2N_{4\a }q_j^{3\a}\Big\}\\
& + \frac{\bfc \|\galpha\|}{\tilde{d} } q_{(n)}.
\end{split}
\]
Applying the Fubini theorem and the formula for integration by  parts, we can write
\[
\begin{split}
\CM_r(q)& := \int_0^{\tilde{d}/q_{(n)}}n_r(\xi) \frac{|b_2(\xi q)|}{\xi}d\xi 
\qquad (\text{with}\quad n_r(\xi):= e^{-(a_0-\eta)\xi^\a}\xi^r)  \\
& \leq  2\Big| S(D)\int_0^{\tilde{d}/q_{(n)}} (1-\cos(\xi q D))\frac{ n_r(\xi)}{\xi}d\xi\Big|  
+ 2q\Big| \int_D^{+\infty}S(x)dx
\int_0^{\tilde{d}/q_{(n)}} n_r(\xi)\sin(\xi q x)d\xi \Big| \\
& \leq \mid S(D) \mid q^2 D^2 N_{r+2}+ \CM^{(1)}_r(q)
\end{split}
\]
where
\[
\begin{split}
\CM^{(1)}_r(q) & :=  2q\Big| \int_D^{+\infty}S(x)dx\int_0^{\tilde{d}/q_{(n)}} n_r(\xi)\sin(\xi q x)d\xi \Big| \\
& \leq   2q \int_D^{+\infty} \frac{|S(x)|}{x}dx \int_0^{+\infty} (1-\cos(\xi qx))\mid \frac{d}{d\xi} n_r(\xi) \mid d\xi  \qquad
(\text{from integration by parts}) \\
& \leq   2 \int_D^{+\infty} \frac{|S(x)|}{x} \int_0^{+\infty}(1\wedge \frac{(\xi qx)^2}{2})
\mid \frac{d}{d\xi} n_r(\xi)\mid d\xi \\
& \leq   2 z_r\big\{q^2\int_0^{1/q}x |S(x)| dx + \int_{1/q}^{+\infty} \frac{|S(x)|}{x}dx \big\}  
\\
&
\qquad \qquad \quad \Big( \text{with}\;\;
z_r:= \max \Big \{\int_0^{+\infty} \mid \frac{d}{d\xi} n_r(\xi)\mid d\xi, 
\frac{1}{2} \int_0^{+\infty} \xi^2 \mid \frac{d}{d\xi} n_r(\xi)\mid d\xi \Big \} \Big)\\
& =  2 z_r\big\{q^{\a}H_1(q)+q^{\a}H_3(q)\}.
\end{split}
\]
Then,
\[
\CM_r(q)  \leq q^2 \mid S(D) \mid D^2 N_{r+2}+2 q^{\a}z_r\{H_1(q)+H_3(q)\}
\]
and
\[
\begin{split}
K(F_n,G_{\a}) & \leq \frac{2}{\pi} \sum_{j=1}^n\big\{ \CM_0(q_j)+ 2 \bar{k}_2\CM_{\a}(q_j)
 + (2 \bar{k}_1 N_2 +8  \bar{k}_1 \bar{k}_2 N_{\a+2})q_j^2  \\ 
&+8 \bar{k}_1^2 N_4q_j^{4-\a}+ \frac{4}{5}M^2 N_{2\a }q_j^{2\a}+
\frac{32}{25}M^2N_{4\a }q_j^{3\a}\big\} + \frac{\bfc \|\galpha\|}{\tilde{d} } q_{(n)}.
\end{split}
\]
To complete the proof it suffices to replace the quantities $\CM$ with their 
upper bounds and, next, to recall the definition
of the constants $\bar{B}$.$\quad \fine$

%\appendix

\section{Appendix}\label{Ap}

In this part of the paper we present the proofs of the theorems stated in Section \ref{s2}. For the sake of 
expository clarity, let us recall the common inspiring principles for all of these proofs. First of all, we refer
to representation (\ref{eq5bis}) which, combined with (\ref{eq13}), gives
\begin{equation}\label{A1}
\mid \phi(\xi,t)- \trgal(\xi)\mid \leq E_t(\mid \tilde{\phi}_{\nu_t}(\xi ; Re( \phi_0) )- \trgal(\xi)\mid)+\mid 
Im (\phi_0(\xi)) \mid e^{-t}
\qquad (\xi \in \RE )
\end{equation}
where $\tilde{\phi}_{\nu_t}(\,\cdot \, ; \, Re (\phi_0) )$ is equal to $\tilde{\phi}_n(\cdot)$ 
when $n=\nu_t$, $q_j=|\beta_{j,t}|$
$(j=1,\dots, \nu_t)$ and,
 in the definition of $\tilde \phi_n$, $\phi_0$ is replaced by $Re \phi_0$. Analogously,
\begin{equation}\label{A2}
\mid F(x,t)- G_{\a}(x) \mid \leq E_t(\mid F_{\nu_t}(x; F^*_0)- G_{\a}(x) \mid) + \mid F_0(x)- F^*_0(x)\mid e^{-t} \qquad (x\in \RE)
\end{equation}
where $ F_{\nu_t}(\, \cdot\,;\, F^*_0)$ is obtained from $F_n(\cdot)$ by replacing $n$, $q_j$ and $F_0$ with $\nu_t$, $|\beta_{j,t}|$
and $F^*_0$, respectively.

\noindent
{\bf Proof of Theorem \ref{thm2}.} Apply (\ref{A1}) to write
\[
\chi_\a(F(\cdot,t),G_\a)  \leq E_t( \chi_\a(F_{\nu_t}(\,\cdot\,; F^*_0), G_{\a}))+  e^{-t}\sup_{\xi \in \RE} \mid Im( v_0(\xi ))
\mid
\]
and, next, replace $\chi_\a(F_{\nu_t}(\,\cdot\,; F^*_0), G_{\a})$ with its upper bound stated in Proposition \ref{prop1}.$\quad \fine$
\vspace{0.3cm}

\noindent
{\bf Proof of Theorem \ref{thm3}.} Argue as in the previous proof by using the upper bounds obtained in Proposition
\ref{prop2}, instead of the upper bound of Proposition \ref{prop1}. 
Moreover, to evaluate expectations, make use of  the obvious inequality
$E_t(\b_{(\nu_t)}^m) \leq E_t[\sum_{j=1}^{\nu_t}|\b_{j,t}|^m]$ and, then, of
(\ref{eq16}) and (\ref{eq17}).$\quad \fine$
\vspace{0.3cm}

\noindent
{\bf Proof of Theorem \ref{thm4}.} In view of (\ref{A2}), write
\[
K(F(\cdot,t),G_\a)  \leq E_t(K(F_{\nu_t}(\,\cdot\, ; F^*_0), G_{\a}))+ \frac{e^{-t}}{2}\sup_{x \in \RE}\mid F_0(x)+
F_0(-x-0)-1\mid
\]
and replace $ K(F_{\nu_t}(\,\cdot \,; F^*_0), G_{\a})$ with its upper bound determined in Proposition \ref{prop3}. 
Finally use 
(\ref{eq16}) to evaluate expectation. $\quad \fine$

The remaining theorems from \ref{thm5}  to \ref{thm9} can be proved following the same line of reasoning, according to
the scheme: Resort to Proposition \ref{prop4} and to (\ref{eq16}) for Theorem \ref{thm5}. Apply Proposition \ref{prop5}
and (\ref{eq16})-(\ref{eq17}) to prove Theorems \ref{thm6} and \ref{thm7}. Finally, use Proposition \ref{prop6} and 
(\ref{eq16})-(\ref{eq17}) to prove Theorems \ref{thm8} and \ref{thm9}.

It remains to prove Theorem \ref{Thm1}. Its former part is a straightforward consequence of Theorem \ref{thm4}. 
As to the latter, we use the same argument as in the proof of Theorem 1 
in \cite{GabettaRegazziniCLT}, 
based on \cite{FortiniLadelliRegazzini}. Accordingly, for every $t>0$ 
we define
\[
W_t:= (\Lambda_{\nu_t}, \lambda_{1,t}, \dots  \lambda_{\nu_t,t},\delta_0, 
\dots ,\gamma_t,\theta_t, \nu_t, 
U_t(1/2), U_t(1/3), \dots )
\]
where: $\lambda_{j,t}$ stands for a conditional distribution of $|\beta_{j,t}|X_{j,t}^*$,
 given $(\gamma_t,\theta_t, \nu_t)$;
$ \Lambda_{\nu_t}$ is the $\nu_t$-fold convolution of $\lambda_{1,t}, \dots  \lambda_{\nu_t,t}$;
 $\delta_x$ indicates unit
mass at $x$; $U_t(\xi):= \max_{1\leq j \leq \nu_t} \lambda_{j,t}([-\xi , \xi]^c)$. Moreover, the $X_{j,t}^*$ are 
conditionally i.i.d. with common distribution $F_0^*$. To grasp the importance of $W_t$, notice 
that its components represent the essential ingredients of central limit problems. 
As to this fundamental theorem, we refer to Section 16.8 of \cite{FristedtGray}.
The range of  $W_t$ can be seen  as a subset of $S:= \P(\bar{\RE})^{\infty}\times \bar{\G} \times [0,2\pi)^{\infty}\times
\bar{\RE}^{\infty}$, where: $ \bar{\RE}:=[-\infty , +\infty]$; $\P(M)$ indicates the set of all probability measures 
on the Borel class $\CB(M)$ on some metric space $M$; $\bar{\G}$ is a distinguished metrizable compactification of $\G$.
These spaces are endowed with topologies specified in Subsection 3.2 of \cite{GabettaRegazziniCLT}, which make $S$
a separable compact metric space. Now recall that, under the assumption of the latter part of  Theorem \ref{Thm1}, $(
V^*_{t_n}:= \sum_{j=1}^{\nu_{t_n}}|\beta_{j,t_n}|X_{j,t_n}^*)_{n\geq 1}$ must converge in distribution. Next, from Lemma 3
in \cite{GabettaRegazziniCLT}, with slight changes, the sequence of the laws of the vectors $(W_{t_n})_{n\geq 1}$ 
contains a subsequence  $(W_{t_{n'}})_{n'}$ which is weakly convergent to a  probability measure $Q$ supported by 
$ \P(\RE)\times \{ \delta_0 \}^{\infty} \times \bar{\G} \times [0,2\pi)^{\infty}\times \{+\infty\}\times \{0\}^{\infty}$.
At this stage, an application of the Skorokhod representation
theorem (see, e.g., \cite{Billingsley}, \cite{Dudley}), combined both with the properties of the support of $Q$
and  with $(F_1)$, entails the existence of random vectors $ \hat{W}_{t_{n'}}:= (\hat{\Lambda}_{\hat{\nu}_{t_{n'}}}, 
\hat{\lambda}_{1,t_{n'}}, \dots )$ defined on a suitable space $(\hat{\Omega}, \hat{\CF}, \hat{P})$, in such a way that $
{W}_{t_{n'}}$ and $\hat{W}_{t_{n'}}$ have the same law (for every $n'$). Moreover,
\begin{equation}\label{A3}
\begin{split}
\hat{\Lambda}_{\hat{\nu}_{t_{n'}}} & \Rightarrow \hat{\Lambda},  \qquad \hat{\lambda}_{j,t_{n'}}\Rightarrow \delta_0
\quad (j=1,2,\dots )\\
\hat{\nu}_{t_{n'}} & \rightarrow +\infty, \qquad \hat{U}_{t_{n'}}(1/k)\rightarrow 0  \quad (k=1,2,\dots ), \\
\hat{\beta}_{(n')} & := \max\{ |\hat{\beta}_{1,t_{n'}}|, \dots |\hat{\beta}_{\hat{\nu}_{t_{n'}},t_{n'}}| \}\rightarrow 0
\end{split} 
\end{equation}
where the convergence must be understood as pointwise convergence on $\hat{\Omega}$ and $\Rightarrow$ designates weak
convergence of probability measures. From (\ref{A3}) and Theorem 16.24 of \cite{FristedtGray}, 
there is a random L\'evy 
measure $\mu$, symmetric about zero, such that
\begin{equation}\label{A4}
\lim_{n'\rightarrow +\infty}\sum_{j=1}^{\hat{\nu}_{t_{n'}}}\hat{\lambda}_{j,t_{n'}}[x, +\infty)= \lim_{n'\rightarrow +\infty}\sum_{j=1}^{\hat{\nu}_{t_{n'}}}\{1-F_0^*\Big(\frac{x}{|\hat{\beta}_{j,t_{n'}}|}\Big)\}= \mu[x, +\infty)
\end{equation}
holds pointwise on $\hat{\Omega}$ for every  $x>0$. To complete the proof, we assume that $\lim_{x\rightarrow +\infty}
x^{\a}\{1-F^*_0(x)\}= +\infty$ and show that this assumption contradicts (\ref{A4}). Indeed,
the assumption implies that for any $k>0$ there is 
$\epsilon >0$ such that $x^{\a}\{1-F^*_0(x)\}\geq k$ for every $x>1/\epsilon$ and, therefore, 
\[
\begin{split}
\nu_{n,x}& := \sum_{j=1}^{\hat{\nu}_{t_{n'}}} \{1-F_0^*\Big(\frac{x}{|\hat{\beta}_{j,t_{n'}}|}\Big)\} \\
& \geq \frac{k}{x^{\a}}
\J\{\hat{\beta}_{(n')} <x \epsilon \} \sum_{j=1}^{\hat{\nu}_{t_{n'}}} |\hat{\beta}_{j,t_{n'}}|^{\a}\\
& = \frac{k}{x^{\a}}\J\{\hat{\beta}_{(n')} <x \epsilon \} .
\end{split}
\]
Since (\ref{A3})  yields $\hat{\beta}_{(n')} \rightarrow 0$, then ${\limsup}_{n\rightarrow +\infty} \nu_{n,x}\geq 
{k}{x^{-\a}}$, which contradicts (\ref{A4}) in view of the arbitrariness of $k$.

%Whence, the theorem is proved by {\it reductio ad absurdum} argument. 
%\bibliographystyle{asa}
%\bibliography{CLTbib}

\begin{thebibliography}{31}
\newcommand{\enquote}[1]{``#1''}
\expandafter\ifx\csname natexlab\endcsname\relax\def\natexlab#1{#1}\fi

\bibitem[{Billingsley(1999)}]{Billingsley}
Billingsley, P. (1999). \textit{Convergence of Probability Measures}, 2nd. ed. Wiley, New York.

\bibitem[{Bobylev et~al.(2000)Bobylev, Carrillo and
  Gamba}]{BobylevCarrilloGamba2000}
Bobylev, A.~V., Carrillo, J.~A. and Gamba, I.~M. (2000). {On some
  properties of kinetic and hydrodynamic equations for inelastic interactions.}
  \textit{J. Statist. Phys.} {\bf 98} 743--773.

\bibitem[{Bobylev and Cercignani(2002{\natexlab{a}})}]{BobylevCercignani2002a}
Bobylev, A.~V. and Cercignani, C. (2002{\natexlab{a}}). {Moment
  equations for a granular material in a thermal bath.} \textit{J. Statist.
  Phys.} {\bf 106} 547--567.

\bibitem[{Bobylev and Cercignani(2002{\natexlab{b}})}]{BobylevCercignani2002b}
--- (2002{\natexlab{b}}). {Self-similar solutions of the {B}oltzmann
  equation and their applications.} \textit{J. Statist. Phys.} {\bf 106}
  1039--1071.

\bibitem[{Bobylev and Cercignani(2002{\natexlab{c}})}]{BobylevCercignani2002c}
--- (2002{\natexlab{c}}). {Self-similar solutions of the {B}oltzmann
  equation for non-{M}axwell molecules.} \textit{J. Statist. Phys.} {\bf 108}
  713--717.

\bibitem[{Bobylev and Cercignani(2003)}]{BobylevCercignani2003}
--- (2003). {Self-similar asymptotics for the {B}oltzmann equation with
  inelastic and elastic interactions.} \textit{J. Statist. Phys.} {\bf 110}
  333--375.

\bibitem[{Bobylev et~al.(2003)Bobylev, Cercignani and
  Toscani}]{BobylevCercignaniToscani2003}
Bobylev, A.~V., Cercignani, C. and Toscani, G. (2003). {Proof of an
  asymptotic property of self-similar solutions of the {B}oltzmann equation for
  granular materials.} \textit{J. Statist. Phys.} {\bf 111} 403--417.

\bibitem[{Bolley and Carrillo(2007)}]{BolleyCarrillo2007}
Bolley, F. and Carrillo, J.~A. (2007). {Tanaka theorem for inelastic
  {M}axwell models.} \textit{Comm. Math. Phys.} {\bf 276} 287--314.

\bibitem[{Carlen et~al.(2000)Carlen, Carvalho and
  Gabetta}]{CarlenCarvalhoGabetta2000}
Carlen, E.~A., Carvalho, M.~C. and Gabetta, E. (2000). {Central limit
  theorem for {M}axwellian molecules and truncation of the {W}ild expansion.}
  \textit{Comm. Pure Appl. Math.} {\bf 53} 370--397.

\bibitem[{Carrillo et~al.(2000)Carrillo, Cercignani and
  Gamba}]{CarrilloCercignaniaGamba2000}
Carrillo, J.~A., Cercignani, C. and Gamba, I.~M. (2000). {Steady
  states of a {B}oltzmann equation for driven granular media.} \textit{Phys.
  Rev. E}(3) {\bf 62} 7700--7707.

\bibitem[{Chow and Teicher(1997)}]{ChowTeicher1997}
Chow, Y.~S. and Teicher, H. (1997). \textit{Probability Theory}, 3rd ed.
Springer, New York.

\bibitem[{Cram{\'e}r(1962)}]{Cramer1962}
Cram{\'e}r, H. (1962). {On the approximation to a stable probability
  distribution.} In \textit{Studies in Mathematical Analysis and Related
  Topics}. 70--76. Stanford Univ. Press.

\bibitem[{Cram{\'e}r(1963)}]{Cramer1963}
--- (1963). {On asymptotic expansions for sums of independent random
  variables with a limiting stable distribution.} \textit{Sankhy\=a Ser. A} 
{\bf 25}  13-24. \textit{ Addendum, ibid.}  216.


\bibitem[{Dudley(2002)}]{Dudley}
Dudley, R.~M. (2002). \textit{Real Analysis and Probability}, revised reprint. 
Cambridge University Press, Cambridge.

\bibitem[{Ernst and Brito(2002)}]{ErnstBrito2002}
Ernst, M.~H. and Brito, R. (2002). {Scaling solutions of inelastic
  {B}oltzmann equations with over-populated high energy tails.} \textit{J.
  Statist. Phys.} {\bf 109} 407--432. 



\bibitem[{Fortini, Ladelli and Regazzini (1996)}]{FortiniLadelliRegazzini}
Fortini, S., Ladelli, L. and Regazzini, E. (1996). 
{A central limit problem for partially exchangeable random variables. }
%Teor. Veroyatnost. i Primenen.  41  (1996),  no. 2, 353--379;  translation in  
\textit{Theory Probab. Appl.}  {\bf 41}  224--246.

 
\bibitem[{Fristedt and Gray(1997)}]{FristedtGray}
Fristedt, B. and Gray, L. (1997). \textit{A Modern Approach to Probability
  Theory}. Birkh\"auser, Boston.

\bibitem[{Gabetta and Regazzini(2006{\natexlab{a}})}]{GabettaRegazzini2006a}
Gabetta, E. and Regazzini, E. (2006{\natexlab{a}}). {Some new results for
  {M}c{K}ean's graphs with applications to {K}ac's equation.} \textit{J. Stat.
  Phys.} {\bf 125} 947--974.

\bibitem[{Gabetta and Regazzini(2006{\natexlab{b}})}]{GabettaRegazziniCLT}
Gabetta, E. and Regazzini, E. (2006{\natexlab{b}}). {Central limit theorem for the solution of the Kac equation.} 
 I.M.A.T.I.-C.N.R., 26-PV. To appear in \textit{The Annals of Applied 
Probability}.


\bibitem[{Gabetta and Regazzini(2006{\natexlab{c}})}]{GabettaRegazziniCLTspeed}
Gabetta, E. and Regazzini, E. (2006{\natexlab{c}}). {Central limit theorem for the solution of the Kac equation: Speed of approach to equilibrium in weak metrics.} 
 I.M.A.T.I.-C.N.R., 27-PV.


\bibitem[{Gabetta et~al.(1995)Gabetta, Toscani and
  Wennberg}]{GabettaToscaniWennberg1995}
Gabetta, E., Toscani, G. and Wennberg, B. (1995). {Metrics for
  probability distributions and the trend to equilibrium for solutions of the
  {B}oltzmann equation.} \textit{J. Statist. Phys.} {\bf 81} 901--934.

\bibitem[{Galambos(1995)}]{Galambos1995}
Galambos, J. (1995). \textit{Advanced Probability Theory},
2nd ed. Marcel Dekker, New York.

\bibitem[{Goudon et~al.(2002)Goudon, Junca and
  Toscani}]{GoudonJuncaToscani2002}
Goudon, T., Junca, S. and Toscani, G. (2002). {Fourier-based distances
  and {B}erry-{E}sseen like inequalities for smooth densities.}
  \textit{Monatsh. Math.} {\bf 135} 115--136.

\bibitem[{Hall(1981)}]{Hall1981}
Hall, P. (1981). {Two-sided bounds on the rate of convergence to a
  stable law.} \textit{Z. Wahrsch. Verw. Gebiete} {\bf 57} 349--364.

\bibitem[{Ibragimov(1985)}]{Ibragimov1985}
Ibragimov, I.~A. (1985). {Th\'eor\`emes limites pour les marches
  al\'eatoires.} In \textit{\'Ecole d'\'Et\'e de Probabilit\'es de Saint-Flour,
  XIII---1983}. \textit{Lecture Notes in Math.} {\bf 1117} 199--297. Springer,
Berlin.

\bibitem[{Ibragimov and Linnik(1971)}]{IbragimovLinnik1971}
Ibragimov, I.~A. and Linnik, Y.~V. (1971). \textit{Independent and Stationary
  Sequences of Random Variables}. Wolters-Noordhoff Publishing, Groningen. 

\bibitem[{Kac(1956)}]{Kac1956}
Kac, M. (1956). {Foundations of kinetic theory.} In \textit{Proceedings
  of the Third Berkeley Symposium on Mathematical Statistics and Probability,
  1954--1955} {\bf 3}  171--197. University of California
  Press, Berkeley and Los Angeles.

\bibitem[{McKean(1966)}]{McKean1966}
McKean, Jr., H.~P. (1966). {Speed of approach to equilibrium for
  {K}ac's caricature of a {M}axwellian gas.} \textit{Arch. Rational Mech.
  Anal.} {\bf 21} 343--367.

\bibitem[{McKean(1967)}]{McKean1967}
--- (1967). {An exponential formula for solving {B}oltmann's equation
  for a {M}axwellian gas.} \textit{J. Combinatorial Theory} {\bf 2} 358--382.

\bibitem[{Pulvirenti and Toscani(2004)}]{PulvirentiToscani2004}
Pulvirenti, A. and Toscani, G. (2004). {Asymptotic properties of the
  inelastic {K}ac model.} \textit{J. Statist. Phys.} {\bf 114} 1453--1480.

\bibitem[{Rachev(1991)}]{Rachev1991}
Rachev, S.~T. (1991). \textit{Probability Metrics and the Stability of
  Stochastic Models}. Wiley, Chichester.  

\bibitem[{Villani(2006)}]{Villani2006}
Villani, C. (2006). {Mathematics of granular materials.} \textit{J.
  Stat. Phys.} {\bf 124} 781--822.

\bibitem[{Wild(1951)}]{Wild1951}
Wild, E. (1951). {On {B}oltzmann's equation in the kinetic theory of
  gases.} \textit{Proc. Cambridge Philos. Soc.} {\bf 47} 602--609.

\bibitem[{Zolotarev(1986)}]{Zolotarev1986}
Zolotarev, V.~M. (1986). \textit{One-Dimensional Stable Distributions}. 
In \textit{Translations of Mathematical Monographs} {\bf 65} AMS, Providence. 

\end{thebibliography}

%{\bf La biblio \`e da correggere...}

\end{document}